\documentclass[iop]{emulateapj}
\usepackage{times}
\usepackage{amsmath}
\slugcomment{\it Submitted 2019 Aug 21; revised 2019 Nov 18; accepted 2019 Dec 2}
\shorttitle{Hot DOG Outflows}
\shortauthors{Jun et al.}

\def\deg{\ifmmode {^{\circ}}\else {$^\circ$}\fi}
\def\kms{\ifmmode {\rm\,km\,s^{-1}}\else
    ${\rm\,km\,s^{-1}}$\fi}
\def\ergcm2s{\ifmmode {\rm\,erg\,cm^{-2}\,s^{-1}}\else
    ${\rm\,erg\,cm^{-2}\,s^{-1}}$\fi}
\def\ergAcm2s{\ifmmode {\rm\,erg\,cm^{-2}\,s^{-1}\,\AA^{-1}}\else
    ${\rm\,erg\,cm^{-2}\,s^{-1}\,\AA^{-1}}$\fi}
\def\ergs{\ifmmode {\rm\,erg\,s^{-1}}\else
    ${\rm\,erg\,s^{-1}}$\fi}
\def\kmsMpc{\ifmmode {\rm\,km\,s^{-1}\,Mpc^{-1}}\else
    ${\rm\,km\,s^{-1}\,Mpc^{-1}}$\fi}

\begin{document}
\title{Spectral Classification and Ionized Gas Outflows in $z\sim2$ WISE-Selected\\ Hot Dust-Obscured Galaxies}
\author{Hyunsung D. Jun\altaffilmark{1,2}, Roberto J. Assef\altaffilmark{3}, Franz E. Bauer\altaffilmark{4,5,6}, Andrew Blain\altaffilmark{7}, Tanio D{\'{\i}}az-Santos\altaffilmark{3}, Peter R. M. Eisenhardt\altaffilmark{2}, Daniel Stern\altaffilmark{2}, Chao-Wei Tsai\altaffilmark{8}, Edward, L. Wright\altaffilmark{9}, and Jingwen Wu\altaffilmark{8}}
\affiliation{National Astronomical Observatories, Chinese Academy of Sciences, 20A Datun Road, Chaoyang District, Beijing 100012, China}
\altaffiltext{1}{School of Physics, Korea Institute for Advanced Study, 85 Hoegiro, Dongdaemun-gu, Seoul 02455, Korea; hsjun@kias.re.kr}
\altaffiltext{2}{Jet Propulsion Laboratory, California Institute of Technology, 4800 Oak Grove Drive, Pasadena, CA 91109, USA}
\altaffiltext{3}{N\'{u}cleo de Astronom\'{i}a de la Facultad de Ingenier\'{i}a y Ciencias, Universidad Diego Portales, Av. Ej\'{e}rcito Libertador 441, Santiago, Chile}
\altaffiltext{4}{Instituto de Astrof{\'{\i}}sica and Centro de Astroingenier{\'{\i}}a, Facultad de F{\'{i}}sica, Pontificia Universidad Cat{\'{o}}lica de Chile, Casilla 306, Santiago 22, Chile}
\altaffiltext{5}{Millennium Institute of Astrophysics (MAS), Nuncio Monse{\~{n}}or S{\'{o}}tero Sanz 100, Providencia, Santiago, Chile} 
\altaffiltext{6}{Space Science Institute, 4750 Walnut Street, Suite 205, Boulder, CO 80301, USA} 
\altaffiltext{7}{Department of Physics and Astronomy, University of Leicester, LE1 7RH Leicester, UK}
\altaffiltext{8}{National Astronomical Observatories, Chinese Academy of Sciences, 20A Datun Road, Chaoyang District, Beijing 100012, China}
\altaffiltext{9}{Department of Physics and Astronomy, University of California, Los Angeles, CA 90095, USA}

\begin{abstract}
We present VLT/XSHOOTER rest-frame UV-optical spectra of 10 Hot Dust-Obscured Galaxies (Hot DOGs) at $z\sim2$ to investigate AGN diagnostics and to assess the presence and effect of ionized gas outflows. Most Hot DOGs in this sample are narrow-line dominated AGN (type 1.8 or higher), and have higher Balmer decrements than typical type 2 quasars. Almost all (8/9) sources show evidence for ionized gas outflows in the form of broad and blueshifted [\ion{O}{3}] profiles, and some sources have such profiles in H$\alpha$ (5/7) or [\ion{O}{2}] (3/6). Combined with the literature, these results support additional sources of obscuration beyond the simple torus invoked by AGN unification models. Outflow rates derived from the broad [\ion{O}{3}] line ($\rm \gtrsim10^{3}\,M_{\odot}\,yr^{-1}$) are greater than the black hole accretion and star formation rates, with feedback efficiencies ($\sim0.1-1\%$) consistent with negative feedback to the host galaxy's star formation in merger-driven quasar activity scenarios. We find the broad emission lines in luminous, obscured quasars are often better explained by outflows within the narrow line region, and caution that black hole mass estimates for such sources in the literature may have substantial uncertainty. Regardless, we find lower bounds on the Eddington ratio for Hot DOGs near unity.
\end{abstract}


\section{Introduction}
According to the triggering and evolution scenario for Active Galactic Nuclei (AGN) from gas rich galaxy mergers (e.g., \citealt{Hop08}; \citealt{Hic09}), the most vigorous accretion in AGN occurs in between the dusty star-forming phase of the merger and the unobscured AGN phase. Gas inflowing towards the merging center is thought to trigger starbursts in a dusty environment, but then the onset of AGN activity introduces negative feedback by heating and pushing out the surrounding gas and dust, disrupting star formation (e.g., \citealt{Sil98}; \citealt{Dim05}; \citealt{Cro06}; \citealt{Fab12}). If there is effective radiative or kinetic energy transferred into the interstellar medium (ISM), the AGN host will become more transparent, until the system becomes quiescent in both AGN and star formation activities.

The interaction between the accreting black hole (BH) and its environment in this scenario results in distinctive phases in BH-galaxy coevolution, and corresponding types of observed galaxies related to each phase. Space-based infrared (IR) missions such as the {\it Infrared Astronomical Satellite} \citep{Neu84}, {\it Spitzer} \citep{Wer04}, {\it Akari} \citep{Mur07} and the {\it Wide-field Infrared Survey Explorer} ({\it WISE}, \citealt{Wri10}) have discovered IR-luminous galaxies over a wide range of luminosity and reddening, sharing observed properties consistent with model predictions. Among the ultraluminous IR galaxies (ULIRGs, $L_{\rm IR}>10^{12} L_{\odot}$, \citealt{San88}), dust-obscured galaxies (DOGs, $F_{\rm \nu, 24\mu m}\rm > 0.3mJy$, $F_{\rm \nu, 24\mu m}/F_{\nu, R\rm-band} \ge 1000$, \citealt{Dey08}) show large amounts of IR emission from dust heated by a mixed contribution of AGN and star formation activities at $z\sim2$, as opposed to starburst dominance found in submillimeter galaxies (SMGs, \citealt{Bla02}; \citealt{Cha05}) at similar luminosity and redshift. Within the merger-triggered AGN evolution scenario in ULIRGs, SMGs are thought to represent an early phase of gas-rich merger-driven starburst, followed by AGN activity taking place in DOGs and clearing out of the AGN surroundings (e.g., \citealt{Dey08}; \citealt{Nar10}). This so-called blowout phase should be observed as AGN outflows in DOGs, as they become transparent and evolve into optically luminous quasars.

With the advent of the {\it WISE} space mission, a similar selection to DOGs was made possible using just the mid-IR flux and color. Objects with $W3$ (12$\mu$m) or $W4$ (22$\mu$m) fluxes an order of magnitude brighter than DOGs, red mid-IR colors, and a faint $W1$ (3.4$\mu$m) flux limit, were discovered and named $W1W2$-dropouts, or Hot DOGs (\citealt{Eis12}; \citealt{Wu12}). Hot DOGs show comparably red $F_{\rm \nu, 24\mu m}/F_{\nu, R\rm-band}$ ratios to DOGs (e.g., \citealt{Wu12}), but higher IR luminosities, pushing into the HyLIRG ($L_{\rm IR}>10^{13} L_{\odot}$, \citealt{San96}) or even ELIRG ($L_{\rm IR}>10^{14} L_{\odot}$, \citealt{Tsa15}) regimes. 

The spectral energy distributions (SEDs) of Hot DOGs are peaked in the mid-IR from dust-heated emission, which together with the extreme IR luminosity suggests the presence and dominance of obscured AGN activity over star formation (e.g., \citealt{Wu12}; \citealt{Ste14}; \citealt{Ass15}; \citealt{Tsa15}). Near-IR through sub-mm imaging observations have also identified their $\sim$arcmin-scale environment to be overdense (e.g., \citealt{Jon14}; \citealt{Ass15}), consistent with a luminous, obscured AGN triggered by mergers of galaxies (e.g., \citealt{Fan16}; \citealt{Ric17b}; \citealt{Dia18}, but see also, \citealt{Far17}). Furthermore, these obscured AGN are nearly as numerous as unobscured AGN at comparably high luminosity (e.g., \citealt{Ass15}; \citealt{Ban15}; \citealt{Lac15}). Hot DOGs are therefore good testbeds to search for strong feedback in action in the most luminous, obscured, and potentially merger-driven AGN. 

Previous studies have reported ionized gas outflows traced by broad and blueshifted motion of the [\ion{O}{3}] emission line not only in $z\lesssim1$ type 2 AGN (e.g., \citealt{Cre00}; \citealt{Gre05a}; \citealt{Vil11}; \citealt{Liu13}; \citealt{Mul13}; \citealt{Zak14}; \citealt{Bae14}; \citealt{Woo16}; \citealt{Dip18}), but also in highly obscured and luminous $z\sim$1--3 AGN where feeding and feedback are expected and observed to be stronger (e.g., \citealt{Gre14}; \citealt{Bru15}; \citealt{Cre15}; \citealt{Per15}; \citealt{Zak16}; \citealt{Leu17}; \citealt{Tob17}; \citealt{Per19}; \citealt{Tem19}). 

The diversity of reddening in AGN, however, is often explained by a mixture of evolutionary and orientation effects (e.g., \citealt{Jun13}; \citealt{She14}; \citealt{Gli18}). Conventionally, the type 1/2 classification for AGN spectra (e.g., \citealt{Ost81}) was interpreted to be an orientation effect of the central obscuring structure to our line-of sight (e.g., \citealt{Ant93}; \citealt{Urr95}), although there have been substantial modifications in response to subsequent observations (e.g., \citealt{Hon07}; \citealt{Nen08}). According to the simplest orientation model, outflows in obscured, type 2 AGN will appear weaker due to a smaller line-of-sight motion of the outflowing material viewed edge-on. 

To better assess the importance of obscured AGN feeding and feedback, as well as to provide AGN diagnostics to differentiate between the simplest form of obscuration models, and to help constrain the spatial extent of dust, we obtained rest-frame UV--optical spectra for 10 Hot DOGs with VLT/XSHOOTER. 
In this paper, we describe our sample and data (\S2), analyze the spectra (\S3), and present the redshifts, line ratios, and the derived accretion and outflow quantities (\S4). A flat, $\Lambda$CDM cosmology with $H_{0}=\mathrm{70\,km\,s^{-1}\,Mpc^{-1}}$, $\Omega_{m}=0.3$, and $\Omega_{\Lambda}=0.7$ (e.g., \citealt{Im97}) is used throughout. 

\section{Data}
\subsection{Sample description and observations}
\begin{figure}
\centering
\includegraphics[scale=.95]{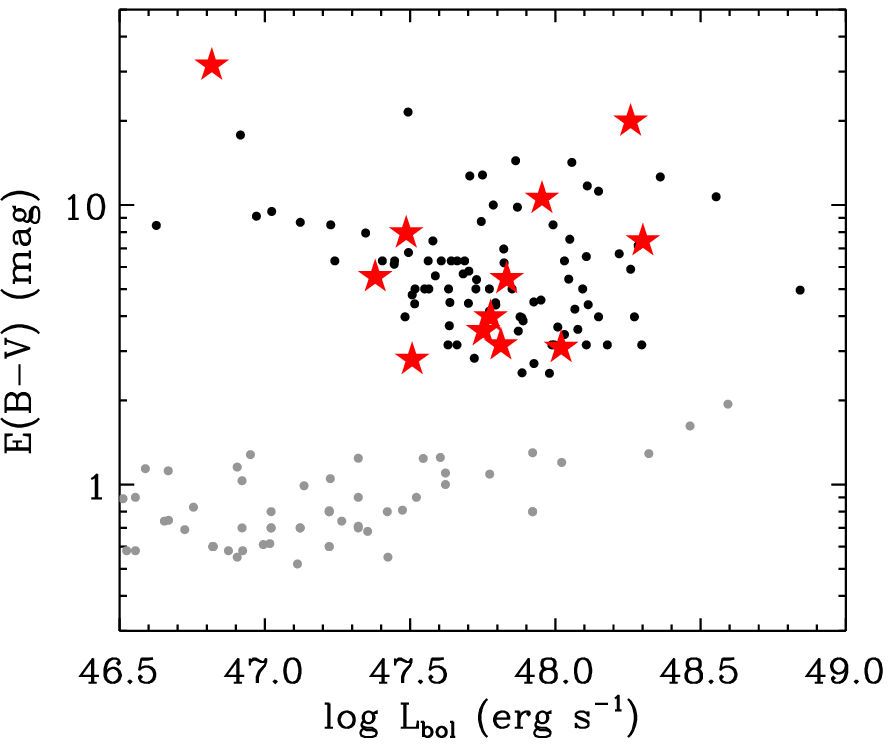}
\caption{Broad-band SED fit-based extinction--bolometric luminosity diagram for Hot DOGs (\citealt{Ass15}, black dots). The XSHOOTER sample is shown by red stars. Red quasars and heavily reddened quasars  (\citealt{Urr12}; \citealt{Ban12}; \citealt{Ban13}; \citealt{Ban15}; \citealt{Kim18}; \citealt{Tem19}, gray dots) are shown for comparison.} 
\end{figure}

\begin{deluxetable*}{cccccccccc}
\tablecolumns{10}
\tabletypesize{\scriptsize}
\tablecaption{Summary of targets}
\tablehead{
\colhead{Name} & \colhead{Coordinates} & \colhead{$z$} & \colhead{flag} & \colhead{$L_{\rm 5100}$} & \colhead{$E(B-V)$} & \colhead{$W1$} & \colhead{$t_{\rm exp}$} & \colhead{Instrument} & \colhead{Wavelength}\\
\colhead{} & \colhead{(J2000)} & \colhead{} & \colhead{} & \colhead{(10$^{46}$\ergs)} & \colhead{(mag)} & \colhead{(mag)} & \colhead{(mag)} & \colhead{(min)} & \colhead{}}
\startdata
W0114--0812 & J011420.48--081243.7 & 2.1037 $\pm$ 0.0002 & A & 6.46 & 4.0  & 17.4 &  50 & 1 & 0.3--2.5$\mu$m\\
W0126--0529 & J012611.96--052909.6 & 0.8301 $\pm$ 0.0001 & A & 0.71 & 31.6 & 17.6 &  50 & 1 & 0.3--2.5$\mu$m\\
W0147--0923 & J014747.58--092350.8 & 2.2535 $\pm$ 0.0006 & A & 19.6 & 20.0 & $>$18.6 & 100 & 1 & 0.3--2.5$\mu$m\\
W0226+0514 & J022646.87+051422.6 & 2.3613 $\pm$ 0.0005 & A & 9.71 & 10.6 & 17.7 & 100 & 1 & 0.3--2.5$\mu$m\\
W1103--1826 & J110330.08--182606.2 & 2.5069 $\pm$ 0.0006 & A & 7.34 & 5.5 &  17.7 & 100 & 1 & 0.3--2.5$\mu$m\\
W1136+4236 & J113634.29+423602.8 & 2.4077 $\pm$ 0.0002 & A & 3.47 & 2.8 & 18.2 & 30, 18 & 2 & $H$, $K$\\
W1719+0446 & J171946.63+044635.2 & 2.5498 $\pm$ 0.0012 & A & 11.3 & 3.1 & 17.6 &  50 & 1 & 0.3--2.5$\mu$m\\
W2016--0041 & J201650.30--004109.0 & 2.6121 $\pm$ 0.0005 & B & 7.01 & 3.2 & 17.9 &  50 & 1 & 0.3--2.5$\mu$m\\
W2026+0716 & J202615.22+071624.2 & 2.5695 $\pm$ 0.0007 & A & 6.09 & 3.6 & 17.6 &  50 & 1& 0.3--2.5$\mu$m\\
W2042--3245 & J204249.28--324517.9 & 2.9583 $\pm$ 0.0058 & B & 21.6 & 7.4 & $>$18.5 &  50 & 1 & 0.3--2.5$\mu$m\\
W2136--1631 & J213655.73--163137.8 & 1.6587 $\pm$ 0.0000 & A & 2.59 & 5.5 & 17.8 &  50 & 1 & 0.3--2.5$\mu$m\\
W2216+0723 & J221619.09+072353.3 & 1.6861 $\pm$ 0.0004 & A & 3.31 & 7.9 & 17.3 & 60 & 3 & $JH$
\enddata
\tablecomments{$z$: systemic redshift (\S4.1), flag: quality of the redshift, $L_{\rm 5100}$: rest-frame 5100\AA\ AGN luminosity, $E(B-V)$: extinction from \citet{Ass15}; Tsai et al. (in preparation), $W1$: AllWISE 3.4$\mu$m Vega magnitude \citep{Wri10}, $t_{\rm exp}$: exposure time,  instrument: VLT/XSHOOTER (1), Keck/MOSFIRE (2), Gemini/FLAMINGOS-2 (3)} 
\end{deluxetable*} 

Our targets, listed in Table 1, were selected to probe rest-frame UV to optical spectral properties of $z\sim2$ Hot DOGs, complementing our observation programs using Keck and Gemini telescopes to investigate BH accretion using the H$\alpha$ line \citep{Wu18}. The XSHOOTER sample consists of 10 galaxies, some with existing redshifts at the time of writing the observation proposal (6/10). The remainder (4/10) were targeted with the added goal of securing a redshift. The color selection for heavily obscured, $W1W2$-dropout sources naturally prefers the rise in the near to mid-IR SED by dust-heated emission from the AGN located in the $W3$-band or longer wavelengths, corresponding to $z\gtrsim2$ (see also, \citealt{Ric17a} for biases against $z\lesssim1$). 

To illustrate where our targets lie in obscuration and luminosity with respect to other populations of luminous, obscured AGN, Figure 1 plots AGN extinction derived from photometric SED fitting, against AGN luminosity.\footnote{Throughout, a bolometric correction of 9.26 (\citealt{Ric06}; \citealt{She11}) is used to convert the extinction-corrected 5100\AA\ luminosity ($L_{5100}$).} The XSHOOTER target samples the $E(B-V)$--$L_{\rm bol}$ space similar to the optical spectroscopic ``Full Sample $W1W2$-dropouts'', in \citet{Ass15}, spanning the highest luminosity and extinction values. Hot DOGs have $E(B-V)$ values of a few to tens of magnitudes, and are up to an order of magnitude more obscured than nearly as luminous, heavily reddened type 1 quasars (\citealt{Ban12}; \citealt{Ban13}; \citealt{Ban15}; \citealt{Tem19}), red type 1 quasars \citep{Gli12}, or extremely red type 1/2 quasars \citep{Ros15} with $E(B-V)\lesssim$1.5 mag. Estimating $E(B-V)$ values for obscured AGN can be quite sensitive to the method. Our values are derived from the photometric SED fitting as described in \citet{Ass15}. When compared to the $E(B-V)$ values from Tsai et al. (in preparation) measured by comparing modeled-to-observed WISE $W2$ flux ratios, the latter $E(B-V)$ values are smaller by about a factor of two, but do not change the main results of this work.

XSHOOTER \citep{Ver11} on VLT was used to collect simultaneous 0.3--2.5$\mu$m spectra of the targets, cross-dispersed along its three arms (UVB, VIS, and NIR). Due to the wide spectral coverage, XSHOOTER is capable of extracting multiple emission lines across the observed spectrum, and obtaining a secure redshift. Slit widths of 1$\arcsec$, 1.2$\arcsec$, and 1.2$\arcsec$ were adopted for the UBV, VIS, and the NIR arms, yielding spectral resolutions of  $R=$\,4290, 3360, and 3900, respectively. The data were taken under program 095.B-0507 in April through September 2015. Each target was observed for either five or ten frames (449s in UBV, 600s in VIS, 514s in NIR, per frame) depending on its $H$-band magnitude \citep{Ass15}. We used an ABBA nodding mode. Calibration stars were observed for each target, as well as arcs for spatial and wavelength calibrations. Typical airmasses during the observations were 1--1.2, with seeing around 1$\arcsec$, and precipitable water vapor of 2--3mm. The observations are summarized in Table 1. In addition to the XSHOOTER spectra, we include previously published Keck/MOSFIRE $H$- and $K$-band spectra ($R\sim3600$) for WISE J113634.29+423602.8, and Gemini/FLAMINGOS-2 $JH$-band ($R\sim1000$) spectra for WISE J221609.09+072353.3 from \citet{Wu18}. 

Supplementing the spectra, we compiled rest-frame optical to mid-IR photometry-based measurements of $L_{\rm bol}$ from \citet{Ass15}, estimated based on the best-fit AGN template after removing host contamination and correcting for obscuration. For unobscured AGN, the monochromatic 5100\AA\ bolometric correction has been widely used to estimate bolometric luminosities from a limited number of photometric bands (e.g., \citealt{Ric06}). In the case of extremely obscured AGN such as Hot DOGs, one could either i) make a reddening correction from the photometric SED and apply a monochromatic bolometric correction (e.g., 5100\AA\ in \citealt{Ass15}), ii) directly integrate along the observed wavelengths without a bolometric correction (e.g., rest-frame UV to far-IR in \citealt{Tsa15}), or iii) make a compromised integration of the best-fit model (e.g., \citealt{Zak16}). The first (i) provides a simple bolometric correction without requiring continuous coverage of the SED, while the second two (ii, iii) allow estimates based on the simple shape of the SED dominated by the mid-IR AGN torus emission: we adopt the \citet{Ass15} formalism (i) for a fair comparison to other studies. 

\begin{figure*}
\centering
\includegraphics[scale=.95]{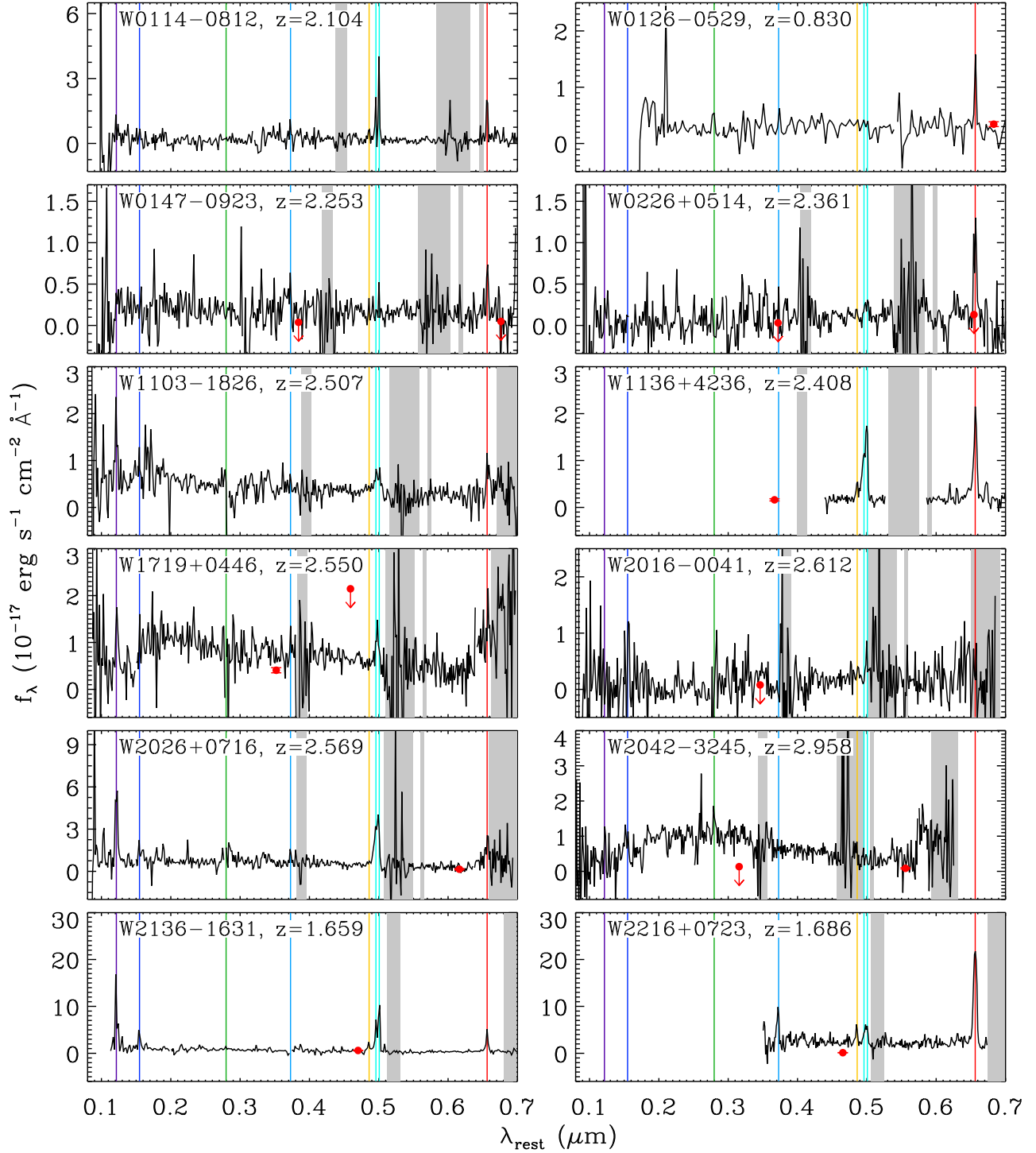}
\caption{The reduced and extracted rest-frame spectra of the sample (black), binned down to $R$=100, 100, 800 for the UBV, VIS, and NIR arms for the sake of display. Photometric data points (red) are from \citet{Ass15}. Wavelengths with strong telluric absorption or sky background are shaded (gray), and noticeable lines (Ly$\alpha$, \ion{C}{4}, \ion{Mg}{2}, [\ion{O}{2}]$\lambda$3727, H$\beta$, [\ion{O}{3}]$\lambda$4959/5007, H$\alpha$) are marked with color-coded vertical lines in order of increasing wavelength. Discontinuities in the spectra are from different arm data.}
\end{figure*}

\subsection{Data reduction}
The ESO Science Archive Facility provides phase-3 data products, reduced with the Reflex pipeline \citep{Fre13}. The reduction steps include bias, dark, flat corrections, wavelength and spatial scale calibrations, flux calibration, combination of frames with cosmic ray rejection, trace and 1-D extraction. In addition to these standard steps, we performed a telluric absorption correction using Molecfit (\citealt{Kau15}; \citealt{Sme15}). Using the standard stars selected close to the target in altitude and azimuth, and observed before or after the science target with the same instrument configuration, we obtained the telluric correction. Visual inspection of the pipeline-processed spectra showed some strongly outlying pixel values outside the telluric bands. We removed these using an absolute S/N threshold of 30--300 times the median S/N, depending on target, but with an uncertainty still less than five times the median noise. These cuts removed 0.15\% of the data, on average, while keeping emission line profiles unchanged.
 
Our data, with 5 or 10 frames, suffers from an insufficient number of frames to fill multiples of four, the number used for a single nodding sequence by the standard ABBA nodding mode reduction pipeline. We therefore use the phase-3 data which only keeps 4/5 or 8/10 frames due to the pipeline limitations. Assuming that object photon noise follows a Poisson distribution, this is an 11\% hit on S/N. When we used 8 frames or twice the nodding sequence, we averaged the flux and propagated the errors from each sequence. When compiling the reduced data coming from adjacent arms, overlapping data near the dichroics (0.560$\mu$m and 1.024$\mu$m) were trimmed to where both sides of the arms have comparable S/N, by constraining $\lambda_{\rm UBV}<0.565\mu$m, $0.555<\lambda_{\rm VIS}<1.03\mu$m, and $1.02<\lambda_{\rm NIR}<2.40\mu$m. Lastly, we applied Galactic extinction corrections assuming $R_{V}=3.1$ and $E(B-V)$ values from the \citet{Sch11} extinction map.

In Figure 2, we plot the reduced, rest-frame 800--7000\AA\ spectra with available photometric fluxes and 2-$\sigma$ upper limits from \citet{Ass15}. Apart from the two spectra from \citet{Wu18} missing the rest-UV wavelengths, the spectra cover major UV--optical emission lines. The average and 1-$\sigma$ scatter of the continuum S/N\footnote{Throughout, we measure the continuum S/N per resolution element, and the line S/N over wavelengths within $\pm$ FWHM from the line center, except [\ion{S}{2}] where we use the flux and uncertainty of the combined doublet.} are $0.9 \pm 1.1$. This is not a critical issue as we focus on the brighter lines, and we apply S/N cuts when analyzing the lines in the forthcoming sections. We binned the spectra matched to the spectral resolution of each arm for analysis.

\section{Analysis}
Taking advantage of the simultaneous, wide spectral coverage of XSHOOTER, we fit the rest-frame 1150--2000\AA, 2000--3500\AA, and 3500--7000\AA\ regions, each containing multiple key emission lines (e.g., Ly$\alpha$/\ion{C}{4}, \ion{Mg}{2}, Balmer and [\ion{O}{3}] lines, respectively). Regions of strong telluric correction (e.g., 1.36--1.41, 1.81--1.96, and 2--2.02$\mu$m) or high sky background (2.4$\mu$m and longer) are masked from the fits. We split the fitting regions into three to account for the shapes of the i) thermal UV bump from the accretion disk, ii) Balmer continuum, iii) broad absorption features, iv) internal extinction and v) host galaxy contamination that can complicate either fitting a simple power-law continuum through the entire wavelength range, or separately constraining the features at low S/N. We used the IDL-based package MPFIT \citep{Mar09} to perform iterative least $\chi^{2}$ fitting with a set of initial parameter values determined from the relative strengths of the features. The fit was improved by applying a 2.5-$\sigma$ Gaussian clipping \citep{Wu18}, and masking some strong outlying values still left on the spectra (\S2.2). Detailed prescriptions are described below, and Table 2 summarizes the broad/narrow nature of the selected lines and the independent/dependent parameters.

\begingroup
\setlength{\tabcolsep}{2pt}
\begin{deluxetable}{cccccc}
\tablecolumns{6}
\tabletypesize{\scriptsize}
\tablecaption{Fitting configuration}
\tablewidth{0.5\textwidth}
\tablehead{
\colhead{Line} & \colhead{Gaussians} & \colhead{Center} & \colhead{FWHM} & \colhead{Flux} & \colhead{Flux Ratio}}
\startdata
Ly$\alpha\lambda$1215 & N+B & 1,2 & 9,10 & 17,18 & ...\\
\ion{N}{5}$\lambda$1240 & N+B & 1,2 & 9,10 & 19,20 & ...\\
\ion{C}{4}$\lambda$1549 & N+B & 1,2 & 9,10 & 21,22 & ...\\
\ion{Si}{3}]$\lambda$1892 & N+B & 1,2 & 9,10 & 23,24 & ...\\
\ion{C}{3}]$\lambda$1908 & N+B & 1,2 & 9,10 & 25,26 & ...\\
\ion{Mg}{2}$\lambda$2798 & N+B & 3,4 & 11,12 & 27,28 & ...\\
$[$\ion{Ne}{5}]$\lambda$3426 & N+B & 3,4 & 11,12 & 29,30 & ...\\
$[$\ion{O}{2}]$\lambda$3727 & N+B & 5,6 & 13,14 & 31,32 & ...\\
H$\gamma\lambda$4340 & N+B & 5,6 & 13,14 & 33,34 & ...\\
$[$\ion{O}{3}]$\lambda$4363 & N & 5 & 13 & 35 & ...\\
H$\beta\lambda$4861 & N+B & 5,6 & 13,14 & 36,37 & ...\\
$[$\ion{O}{3}]$\lambda$4959/5007 & N+B & 7,8 & 15,16 & 38,39 & 2.98\\
$[$\ion{O}{1}]$\lambda$6300 & N & 5 & 13 & 40 & ...\\
$[$\ion{N}{2}]$\lambda$6548/6583 & N & 5 & 13 & 41 & 2.96\\
H$\alpha\lambda$6563 & N+B & 5,6 & 13,14 & 42,43 & ...\\
$[$\ion{S}{2}]$\lambda$6716/6731 & N & 5 & 13 & 44,45 & Free
\enddata
\tablecomments{For each line, N and B represent narrow{ (FWHM\,$\le$\,1200\kms)} and broad (FWHM\,$\le$\,10,000\kms) Gaussian components. The Ly$\alpha$ through [\ion{Ne}{5}] lines are treated with a broad component only when S/N$<$5. The values among the Center, FWHM, and Flux columns represent independent fitting parameters such that identical values indicate they are tied to each other. The Flux Ratio column denotes whether the doublets are treated with a fixed or free line ratio, and the value is the line ratio of the stronger doublet to the weaker.} 
\end{deluxetable} 
\endgroup

For the rest-frame 3500--7000\AA\ fit, we assumed a power-law continuum and Gaussian emission lines for [\ion{O}{2}]$\lambda$3727, H$\gamma$, [\ion{O}{3}]$\lambda$4363, H$\beta$, [\ion{O}{3}]$\lambda$4959/5007, [\ion{O}{1}]$\lambda$6300, [\ion{N}{2}]$\lambda$6548/6583, H$\alpha$, and [\ion{S}{2}]$\lambda$6716/6731. Traditionally, Hydrogen Balmer lines in AGN are fit by both broad and narrow components with the dividing line between these components at FWHM\,=\,1000\kms, while forbidden transitions are treated as narrow lines only. However, this standard approach is inappropriate for asymmetric, blueshifted and broadened forbidden line profiles, as observed in some [\ion{O}{3}] doublets, and even for some [\ion{O}{2}] and Hydrogen Balmer lines in type 2 quasars (e.g., \citealt{Zak14}; \citealt{Kan17}; \citealt{Wu18}). These profiles are often broad, but instead of being broadened by the motions of the broad-line region clouds, they are thought to be broadened by emission from ionized gas outflowing from the AGN at $\sim10^{3}\,\kms$ (\citealt{Vei94}; \citealt{Cre00}). 

Independent of the origin of the broad-line components, we fit a single narrow (FWHM\,$\le$\,1200\kms) and a single broad (FWHM\,$\le$\,10,000\kms) Gaussian to the Balmer lines (H$\alpha$, H$\beta$, H$\gamma$) and [\ion{O}{2}], [\ion{O}{3}] doublets\footnote{We loosen the narrow and broad line width constraints as there are widths converging around 1000--1200\kms\ (Table 3) or the fit converges when the broad component is allowed to be narrower than 1000\kms. When both components and the combined profile have FWHM$<$1200\kms\ we assume the total profile to be narrow.}.
To benefit from multiple lines being covered within the fitting range, we fix the line centers and widths by a single redshift and FWHM for all the narrow components and likewise for the broad Hydrogen lines and [\ion{O}{2}] with undetermined origin (see \S4.5 for the interpretation), except for the narrow/broad [\ion{O}{3}] components where we further use separate redshifts and FWHMs to account for outflows. Exceptions are W2016--0041 (H$\beta$/H$\gamma$) where we removed the broad component as the noise around a narrow line was fit without a clear detection. The [\ion{O}{3}]$\lambda$4363, [\ion{O}{1}], [\ion{N}{2}], and [\ion{S}{2}] lines are weak and do not show any signs of deviation from a narrow profile, and thus were fit with single narrow Gaussians.

We note that tying the Balmer line widths is a simplified assumption since broad H$\beta$ line is empirically 10--20\% wider than broad H$\alpha$ (e.g., \citealt{Jun15}). However, jointly constraining the line center and width is appropriate given the S/N of our data. Also, we allowed the narrow and broad line centers to be within $-1000$ to $1000\kms$ and $-2000$ to $1000\kms$ of the systemic redshift (\S4.1), respectively, to account for the blueshift/redshift of the broad component.
We fix the [\ion{O}{3}] doublet ratio to 2.98 (e.g., \citealt{Sto00}; \citealt{Dim07}), the [\ion{N}{2}] doublet ratio to 2.96 (e.g., \citealt{Gre05b}; \citealt{Wu18}), but leave the [\ion{S}{2}] doublet ratio free as it is less blended with other lines and we can use this line ratio to estimate the electron density (see \S4.4).

\begin{figure*}
\centering
\includegraphics[scale=1.]{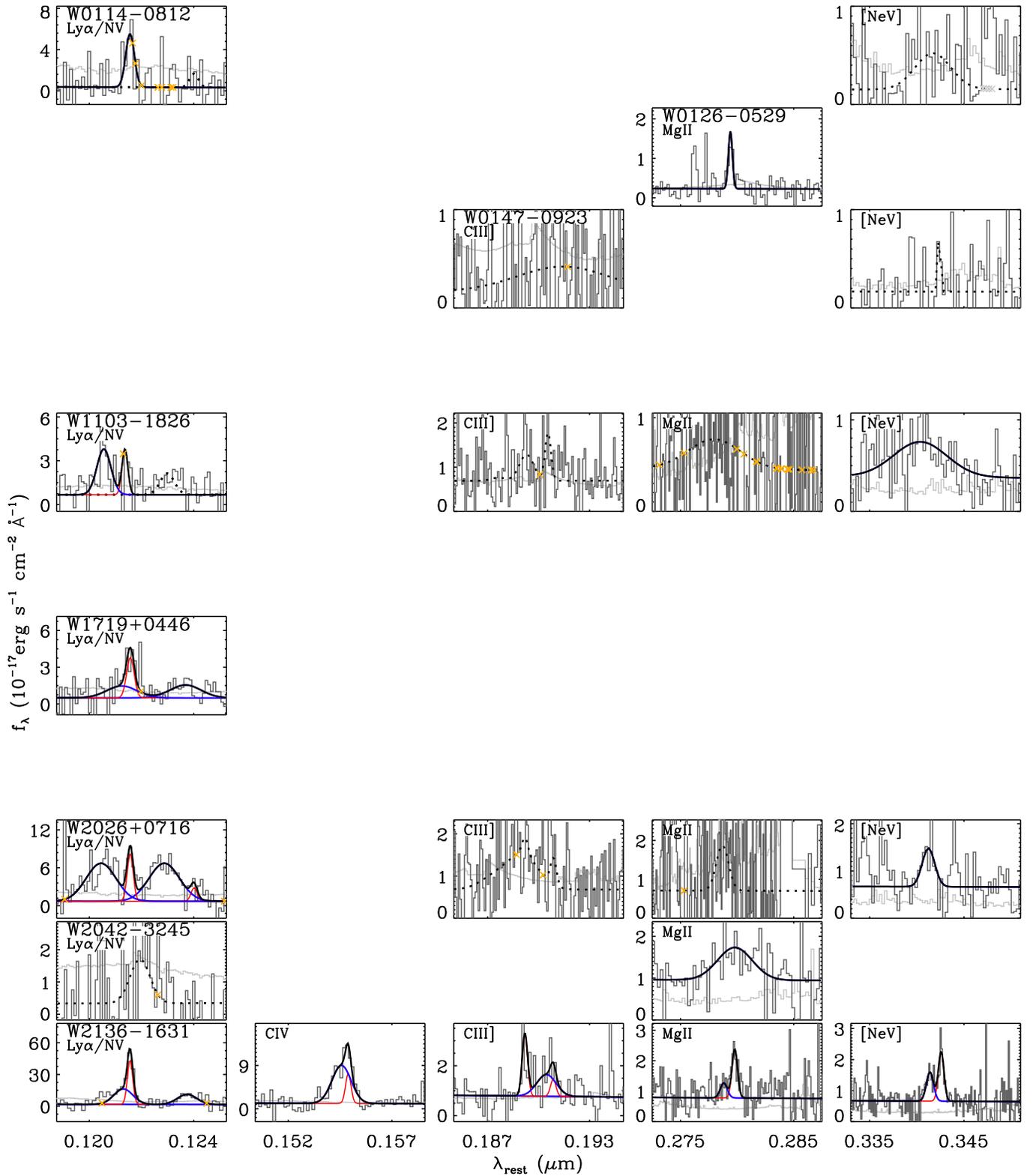}
\caption{Model fit to the rest-frame, resolution-matched spectra of the sample, overplotted on the flux (black histogram) and uncertainty (gray histogram) spectra. The narrow (red) and broad (blue) components are highlighted. Wavelengths masked, with strong telluric absorption, or with high sky background are marked (yellow crosses), and we only show the panels with detected lines (solid for S/N$\ge$3, and dotted for S/N$\ge$2). The wavelength limits for each panel are scaled to show $\pm$\,8000\kms\ from the panel center.}
\end{figure*}

\begin{figure*}
\centering
\includegraphics[scale=.95]{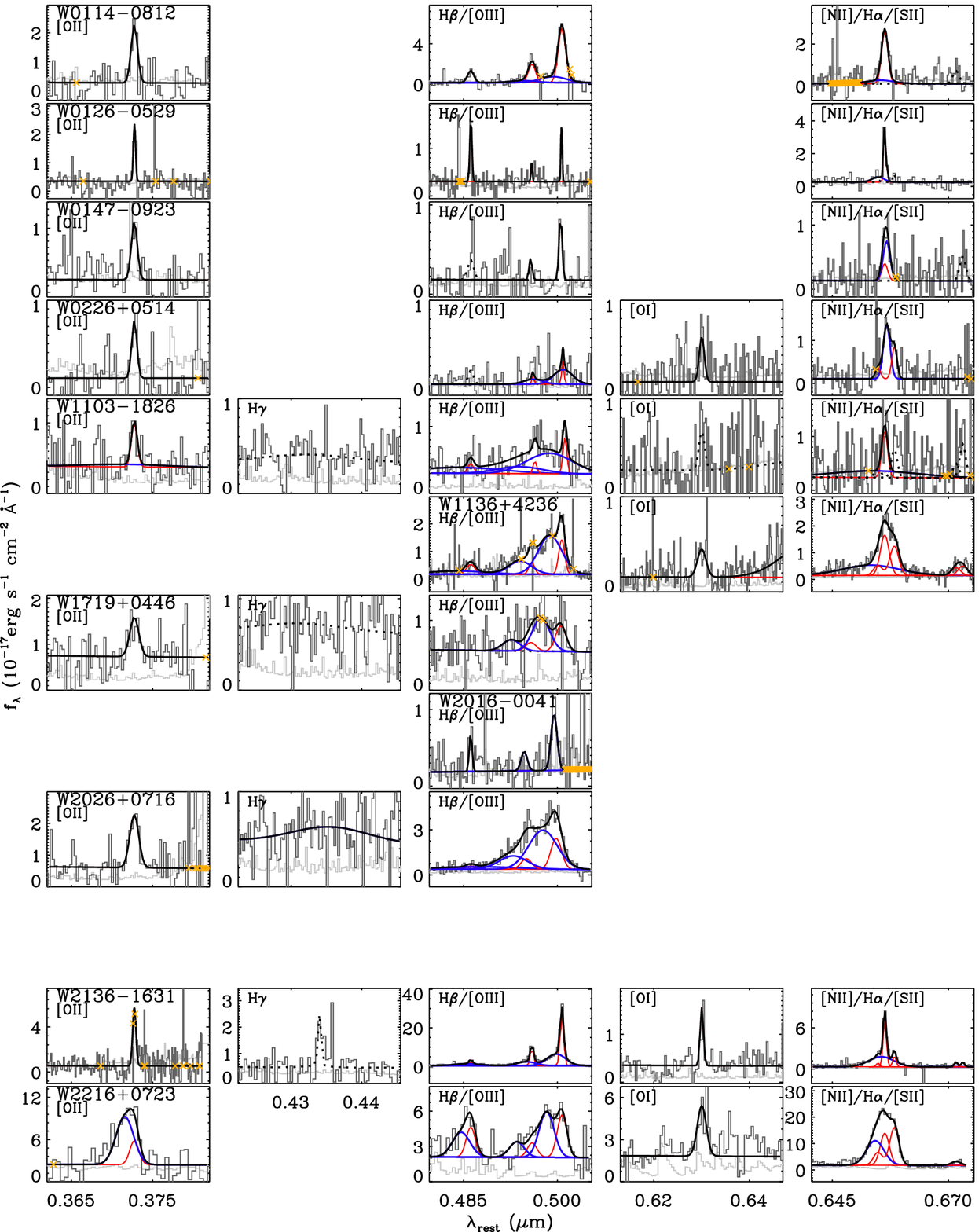}
Continued from Figure 3, for the rest-optical spectra.
\end{figure*}

Next, we fit the rest-frame UV wavelengths, separately in the 1150--2000 and 2000--3500\AA\ windows in order to better fit the continuum, which often shows a downturn at wavelengths shorter than 2000--3000\AA\ (e.g., W2026+0716, W2042--3245, also, e.g., \citealt{Jun13} for quasars in general). We restricted the fitting range of two objects showing significant features in the continuum around \ion{C}{4} (W1719+0446, W2042--3245). We fit a power-law continuum and Gaussian(s) for the Ly$\alpha$, \ion{N}{5}, \ion{C}{4}, \ion{C}{3}] lines within the 1150--2000\AA\ region, and \ion{Mg}{2}, [\ion{Ne}{5}] lines within the 2000--3500\AA\ region. As the targets have a relatively red rest-frame UV-optical color, the rest-UV emission lines are typically weaker than the rest-optical lines. Note that fitting a double Gaussian profile to a low S/N line often overfits a weak, broad component indistinguishable with the noise, which stands out as well-fit lines are consistently narrower. Thus, we only use double Gaussians (i.e., narrow and broad) to fit lines in the 1150--2000 or  2000--3500\AA\ regions with a S/N$\ge$5 detection, and a single Gaussian (FWHM\,$\le$\,10,000\kms) for the others. 

\begingroup
\setlength{\tabcolsep}{3pt}
\begin{deluxetable*}{ccccccccccc}
\tabletypesize{\scriptsize}
\tablecaption{Rest Optical Emission Line Properties}
\tablehead{
\colhead{Name} & \colhead{$\log L$} & \colhead{$\Delta v$} & \colhead{FWHM} & \colhead{$\sigma$} & \colhead{$\log L_{\rm bl}$} & \colhead{$\Delta v_{\rm bl}$} & \colhead{FWHM$_{\rm bl}$} & \colhead{$\log L_{\rm nl}$} & \colhead{$\Delta v_{\rm nl}$} & \colhead{FWHM$_{\rm nl}$}\\
\colhead{} & \colhead{(\ergs)} & \colhead{(\kms)} & \colhead{(\kms)} & \colhead{(\kms)} & \colhead{(\ergs)} & \colhead{(\kms)} & \colhead{(\kms)} & \colhead{(\ergs)} & \colhead{(\kms)} & \colhead{(\kms)}}
\startdata
\sidehead{[\ion{O}{2}]}
W0114--0812 & 43.81 $\pm$ 0.06 & 0 $\pm$ 25 & 735 $\pm$ 52 & 312 $\pm$ 22 & ... $\pm$ ... & ... $\pm$ ... & ... $\pm$ ... & 43.81 $\pm$ 0.06 & 0 $\pm$ 25 & 735 $\pm$ 52\\
W0126--0529 & 42.37 $\pm$ 0.07 & 25 $\pm$ 17 & 250 $\pm$ 27 & 106 $\pm$ 12 & ... $\pm$ ... & ... $\pm$ ... & ... $\pm$ ... & 42.37 $\pm$ 0.07 & 25 $\pm$ 17 & 250 $\pm$ 27\\
W0147--0923 & 43.51 $\pm$ 0.12 & 0 $\pm$ 80 & 686 $\pm$ 123 & 291 $\pm$ 52 & ... $\pm$ ... & ... $\pm$ ... & ... $\pm$ ... & 43.51 $\pm$ 0.12 & 0 $\pm$ 80 & 686 $\pm$ 123\\
W0226+0514 & 43.36 $\pm$ 0.10 & 0 $\pm$ 57 & 599 $\pm$ 87 & 254 $\pm$ 37 & ... $\pm$ ... & ... $\pm$ ... & ... $\pm$ ... & 43.36 $\pm$ 0.10 & 0 $\pm$ 57 & 599 $\pm$ 87\\
W1103--1826 & 43.72 $\pm$ 0.25 & --23 $\pm$ 50 & 724 $\pm$ 104 & 662 $\pm$ 207 & 43.38 $\pm$ 0.52 & --1122 $\pm$ 1130 & 10000 $\pm$ 38 & 43.45 $\pm$ 0.11 & 0 $\pm$ 67 & 676 $\pm$ 104\\
W1719+0446 & 43.79 $\pm$ 0.14 & 0 $\pm$ 138 & 1048 $\pm$ 247 & 445 $\pm$ 105 & ... $\pm$ ... & ... $\pm$ ... & ... $\pm$ ... & 43.79 $\pm$ 0.14 & 0 $\pm$ 138 & 1048 $\pm$ 247\\
W2026+0716 & 44.10 $\pm$ 0.08 & 0 $\pm$ 86 & 1049 $\pm$ 138 & 446 $\pm$ 59 & ... $\pm$ ... & ... $\pm$ ... & ... $\pm$ ... & 44.10 $\pm$ 0.08 & 0 $\pm$ 86 & 1049 $\pm$ 138\\
W2136--1631 & 43.68 $\pm$ 0.05 & --1 $\pm$ 3 & 379 $\pm$ 7 & 161 $\pm$ 3 & ... $\pm$ ... & ... $\pm$ ... & ... $\pm$ ... & 43.68 $\pm$ 0.05 & --1 $\pm$ 3 & 379 $\pm$ 7\\
W2216+0723 & 44.65 $\pm$ 0.11 & --751 $\pm$ 49 & 2306 $\pm$ 375 & 890 $\pm$ 99 & 44.55 $\pm$ 0.11 & --944 $\pm$ 215 & 2056 $\pm$ 299 & 43.96 $\pm$ 0.36 & 0 $\pm$ 65 & 1054 $\pm$ 71\\
\sidehead{H$\beta$}
W0114--0812 & 43.69 $\pm$ 0.06 & 0 $\pm$ 24 & 735 $\pm$ 52 & 312 $\pm$ 22 & ... $\pm$ ... & ... $\pm$ ... & ... $\pm$ ... & 43.69 $\pm$ 0.06 & 0 $\pm$ 24 & 735 $\pm$ 52\\
W0126--0529 & 42.25 $\pm$ 0.08 & 25 $\pm$ 16 & 250 $\pm$ 27 & 106 $\pm$ 12 & ... $\pm$ ... & ... $\pm$ ... & ... $\pm$ ... & 42.25 $\pm$ 0.08 & 25 $\pm$ 16 & 250 $\pm$ 27\\
W1103--1826 & 43.92 $\pm$ 0.10 & --522 $\pm$ 53 & 1025 $\pm$ 115 & 1582 $\pm$ 4 & 43.88 $\pm$ 0.11 & --1122 $\pm$ 1130 & 10000 $\pm$ 38 & 42.93 $\pm$ 0.23 & 0 $\pm$ 67 & 676 $\pm$ 104\\
W1136+4236 & 43.95 $\pm$ 0.06 & --424 $\pm$ 17 & 1259 $\pm$ 41 & 1548 $\pm$ 36 & 43.79 $\pm$ 0.08 & --1360 $\pm$ 167 & 6533 $\pm$ 324 & 43.45 $\pm$ 0.06 & 0 $\pm$ 24 & 1022 $\pm$ 41\\
W2016--0041 & 43.19 $\pm$ 0.16 & 0 $\pm$ 55 & 310 $\pm$ 84 & 131 $\pm$ 36 & ... $\pm$ ... & ... $\pm$ ... & ... $\pm$ ... & 43.19 $\pm$ 0.16 & 0 $\pm$ 55 & 310 $\pm$ 84\\
W2026+0716 & 44.15 $\pm$ 0.15 & 545 $\pm$ 110 & 1880 $\pm$ 278 & 2573 $\pm$ 136 & 44.09 $\pm$ 0.17 & 986 $\pm$ 61 & 8726 $\pm$ 2303 & 43.26 $\pm$ 0.23 & 0 $\pm$ 86 & 1049 $\pm$ 138\\
W2136--1631 & 44.07 $\pm$ 0.04 & --216 $\pm$ 3 & 841 $\pm$ 18 & 963 $\pm$ 14 & 43.97 $\pm$ 0.04 & --334 $\pm$ 42 & 2870 $\pm$ 93 & 43.42 $\pm$ 0.05 & 0 $\pm$ 3 & 379 $\pm$ 7\\
W2216+0723 & 44.36 $\pm$ 0.10 & --582 $\pm$ 59 & 1865 $\pm$ 243 & 870 $\pm$ 83 & 44.15 $\pm$ 0.13 & --944 $\pm$ 215 & 2056 $\pm$ 299 & 43.95 $\pm$ 0.17 & 0 $\pm$ 64 & 1054 $\pm$ 71\\
\sidehead{[\ion{O}{3}]}
W0114--0812 & 44.56 $\pm$ 0.04 & --158 $\pm$ 17 & 889 $\pm$ 30 & 695 $\pm$ 48 & 44.00 $\pm$ 0.12 & --815 $\pm$ 210 & 2920 $\pm$ 335 & 44.41 $\pm$ 0.02 & 20 $\pm$ 20 & 827 $\pm$ 29\\
W0126--0529 & 42.06 $\pm$ 0.13 & --1 $\pm$ 19 & 148 $\pm$ 39 & 63 $\pm$ 17 & ... $\pm$ ... & ... $\pm$ ... & ... $\pm$ ... & 42.06 $\pm$ 0.13 & --1 $\pm$ 19 & 148 $\pm$ 39\\
W0147--0923 & 43.24 $\pm$ 0.09 & --108 $\pm$ 59 & 333 $\pm$ 55 & 142 $\pm$ 23 & ... $\pm$ ... & ... $\pm$ ... & ... $\pm$ ... & 43.24 $\pm$ 0.09 & --108 $\pm$ 59 & 333 $\pm$ 55\\
W0226+0514 & 43.48 $\pm$ 0.15 & 125 $\pm$ 41 & 440 $\pm$ 110 & 603 $\pm$ 35 & 43.40 $\pm$ 0.13 & 145 $\pm$ 248 & 2111 $\pm$ 568 & 42.68 $\pm$ 0.24 & 112 $\pm$ 69 & 244 $\pm$ 104\\
W1103--1826 & 44.21 $\pm$ 0.06 & --891 $\pm$ 74 & 1788 $\pm$ 194 & 1980 $\pm$ 170 & 44.15 $\pm$ 0.09 & --1172 $\pm$ 266 & 5211 $\pm$ 662 & 43.32 $\pm$ 0.09 & 309 $\pm$ 56 & 432 $\pm$ 73\\
W1136+4236 & 44.57 $\pm$ 0.02 & --905 $\pm$ 17 & 2362 $\pm$ 82 & 1116 $\pm$ 36 & 44.48 $\pm$ 0.03 & --1110 $\pm$ 56 & 2662 $\pm$ 96 & 43.83 $\pm$ 0.04 & 8 $\pm$ 25 & 668 $\pm$ 51\\
W1719+0446 & 44.10 $\pm$ 0.10 & --1359 $\pm$ 141 & 1326 $\pm$ 313 & 1151 $\pm$ 123 & 43.94 $\pm$ 0.45 & --1968 $\pm$ 97 & 1988 $\pm$ 412 & 43.61 $\pm$ 0.18 & --82 $\pm$ 184 & 1189 $\pm$ 369\\
W2016--0041 & 43.71 $\pm$ 0.14 & --721 $\pm$ 102 & 691 $\pm$ 192 & 293 $\pm$ 81 & 43.71 $\pm$ 0.28 & --721 $\pm$ 102 & 691 $\pm$ 192 & ... $\pm$ ... & ... $\pm$ ... & ... $\pm$ ...\\
W2026+0716 & 45.01 $\pm$ 0.02 & --1495 $\pm$ 61 & 3000 $\pm$ 107 & 1338 $\pm$ 41 & 44.90 $\pm$ 0.04 & --1785 $\pm$ 85 & 3227 $\pm$ 121 & 44.37 $\pm$ 0.03 & --519 $\pm$ 69 & 1200 $\pm$ 0\\
W2136--1631 & 44.87 $\pm$ 0.01 & --215 $\pm$ 3 & 609 $\pm$ 6 & 613 $\pm$ 7 & 44.61 $\pm$ 0.01 & --515 $\pm$ 21 & 2089 $\pm$ 32 & 44.53 $\pm$ 0.01 & 18 $\pm$ 3 & 387 $\pm$ 4\\
W2216+0723 & 44.55 $\pm$ 0.07 & --882 $\pm$ 47 & 2811 $\pm$ 1096 & 960 $\pm$ 97 & 44.34 $\pm$ 0.10 & --1443 $\pm$ 164 & 1690 $\pm$ 348 & 44.13 $\pm$ 0.12 & 38 $\pm$ 126 & 1099 $\pm$ 214\\
\sidehead{H$\alpha$}
W0114--0812 & 44.25 $\pm$ 0.11 & --60 $\pm$ 37 & 771 $\pm$ 54 & 568 $\pm$ 193 & 43.59 $\pm$ 0.45 & --430 $\pm$ 485 & 3074 $\pm$ 1146 & 44.15 $\pm$ 0.05 & --1 $\pm$ 24 & 735 $\pm$ 52\\
W0126--0529 & 43.03 $\pm$ 0.11 & --188 $\pm$ 130 & 890 $\pm$ 436 & 461 $\pm$ 230 & 42.52 $\pm$ 0.31 & --650 $\pm$ 396 & 1371 $\pm$ 767 & 42.86 $\pm$ 0.06 & 24 $\pm$ 16 & 250 $\pm$ 27\\
W0147--0923 & 43.79 $\pm$ 1.08 & 131 $\pm$ 75 & 808 $\pm$ 216 & 339 $\pm$ 73 & 43.65 $\pm$ 1.05 & 179 $\pm$ 429 & 805 $\pm$ 212 & 43.22 $\pm$ 2.82 & --1 $\pm$ 80 & 686 $\pm$ 123\\
W0226+0514 & 44.00 $\pm$ 0.09 & 188 $\pm$ 63 & 780 $\pm$ 140 & 331 $\pm$ 60 & 44.00 $\pm$ 0.09 & 188 $\pm$ 63 & 780 $\pm$ 140 & ... $\pm$ ... & ... $\pm$ ... & ... $\pm$ ...\\
W1103--1826 & 44.33 $\pm$ 0.14 & --49 $\pm$ 47 & 760 $\pm$ 104 & 902 $\pm$ 59 & 44.16 $\pm$ 0.20 & --1122 $\pm$ 1130 & 10000 $\pm$ 38 & 43.84 $\pm$ 0.12 & 0 $\pm$ 67 & 676 $\pm$ 104\\
W1136+4236 & 44.63 $\pm$ 0.02 & --335 $\pm$ 23 & 1194 $\pm$ 41 & 1401 $\pm$ 23 & 44.42 $\pm$ 0.03 & --1360 $\pm$ 167 & 6533 $\pm$ 324 & 44.21 $\pm$ 0.02 & 0 $\pm$ 23 & 1022 $\pm$ 41\\
W2136--1631 & 44.46 $\pm$ 0.01 & --69 $\pm$ 3 & 467 $\pm$ 7 & 560 $\pm$ 6 & 44.25 $\pm$ 0.02 & --335 $\pm$ 42 & 2870 $\pm$ 93 & 44.05 $\pm$ 0.01 & --1 $\pm$ 4 & 379 $\pm$ 7\\
W2216+0723 & 45.15 $\pm$ 0.06 & --565 $\pm$ 64 & 1791 $\pm$ 221 & 866 $\pm$ 93 & 44.92 $\pm$ 0.09 & --944 $\pm$ 215 & 2056 $\pm$ 299 & 44.75 $\pm$ 0.09 & 0 $\pm$ 65 & 1054 $\pm$ 71
\enddata
\tablecomments{$L$ is the luminosity, $\Delta v$ the offset (luminosity weighted 1st moment) for the total profile or the Gaussian center for each broad/narrow component, with respect to the systemic redshift, FWHM the full width at half maximum, and $\sigma$ the line dispersion (luminosity weighted 2nd moment), respectively. Subscripts ``bl'' and ``nl'' indicate broad line and narrow line, respectively. The properties of the broad and narrow components are denoted with subscripts. We show only the fitted values for detected lines (S/N$\ge$3 from the total profile), and values from a component without any contribution to the model are left blank.}
\end{deluxetable*} 
\endgroup

\begingroup
\setlength{\tabcolsep}{3pt}
\begin{deluxetable*}{ccccccccccc}
\tabletypesize{\scriptsize}
\tablecaption{Rest far-UV Emission Line Properties}
\tablehead{
\colhead{Name} & \colhead{$\log L$} & \colhead{$\Delta v$} & \colhead{FWHM} & \colhead{$\sigma$} & \colhead{$\log L_{\rm bl}$} & \colhead{$\Delta v_{\rm bl}$} & \colhead{FWHM$_{\rm bl}$} & \colhead{$\log L_{\rm nl}$} & \colhead{$\Delta v_{\rm nl}$} & \colhead{FWHM$_{\rm nl}$}\\
\colhead{} & \colhead{(\ergs)} & \colhead{(\kms)} & \colhead{(\kms)} & \colhead{(\kms)} & \colhead{(\ergs)} & \colhead{(\kms)} & \colhead{(\kms)} & \colhead{(\ergs)} & \colhead{(\kms)} & \colhead{(\kms)}}
\startdata
\sidehead{Ly$\alpha$}
W0114--0812 & 43.85 $\pm$ 0.20 & 56 $\pm$ 134 & 960 $\pm$ 309 & 408 $\pm$ 131 & 43.85 $\pm$ 0.20 & 56 $\pm$ 134 & 960 $\pm$ 309 & ... $\pm$ ... & ... $\pm$ ... & ... $\pm$ ...\\
W1103--1826 & 44.18 $\pm$ 0.12 & --1827 $\pm$ 94 & 3120 $\pm$ 1072 & 1082 $\pm$ 103 & 44.02 $\pm$ 0.14 & --2415 $\pm$ 158 & 1571 $\pm$ 368 & 43.64 $\pm$ 0.23 & --421 $\pm$ 109 & 653 $\pm$ 236\\
W1719+0446 & 44.13 $\pm$ 0.19 & --247 $\pm$ 97 & 1002 $\pm$ 279 & 988 $\pm$ 179 & 43.87 $\pm$ 0.30 & --716 $\pm$ 542 & 3371 $\pm$ 1002 & 43.80 $\pm$ 0.20 & 91 $\pm$ 144 & 843 $\pm$ 290\\
W2026+0716 & 44.76 $\pm$ 0.06 & --2119 $\pm$ 74 & 2432 $\pm$ 250 & 1689 $\pm$ 110 & 44.66 $\pm$ 0.07 & --2676 $\pm$ 155 & 3334 $\pm$ 368 & 44.06 $\pm$ 0.13 & 81 $\pm$ 89 & 667 $\pm$ 164\\
W2136--1631 & 44.64 $\pm$ 0.07 & --245 $\pm$ 8 & 727 $\pm$ 71 & 682 $\pm$ 37 & 44.38 $\pm$ 0.10 & --574 $\pm$ 104 & 2102 $\pm$ 189 & 44.30 $\pm$ 0.08 & 58 $\pm$ 27 & 592 $\pm$ 75\\
\sidehead{\ion{N}{5}}
W1719+0446 & 43.91 $\pm$ 0.19 & --716 $\pm$ 541 & 3371 $\pm$ 1002 & 1431 $\pm$ 425 & 43.91 $\pm$ 0.19 & --716 $\pm$ 541 & 3371 $\pm$ 1002 &  ... $\pm$ ... &  ... $\pm$ ... &  ... $\pm$ ...\\
W2026+0716 & 44.70 $\pm$ 0.06 & --2495 $\pm$ 76 & 4406 $\pm$ 521 & 1534 $\pm$ 152 & 44.67 $\pm$ 0.06 & --2677 $\pm$ 155 & 3334 $\pm$ 368 & 43.52 $\pm$ 0.29 & 80 $\pm$ 88 & 667 $\pm$ 164\\
W2136--1631 & 44.21 $\pm$ 0.11 & --576 $\pm$ 104 & 2102 $\pm$ 189 & 893 $\pm$ 80 & 44.21 $\pm$ 0.11 & --576 $\pm$ 104 & 2102 $\pm$ 189 &  ... $\pm$ ... &  ... $\pm$ ... &  ... $\pm$ ...\\
\sidehead{\ion{C}{4}}
W2136--1631 & 44.31 $\pm$ 0.05 & --460 $\pm$ 5 & 1621 $\pm$ 131 & 849 $\pm$ 67 & 44.22 $\pm$ 0.06 & --576 $\pm$ 103 & 2101 $\pm$ 189 & 43.57 $\pm$ 0.12 & 57 $\pm$ 27 & 589 $\pm$ 75\\
\sidehead{\ion{C}{3}]}
W2136--1631 & 43.44 $\pm$ 0.16 & --461 $\pm$ 8 & 1629 $\pm$ 132 & 850 $\pm$ 38 & 43.35 $\pm$ 0.18 & --575 $\pm$ 104 & 2101 $\pm$ 189 & 42.70 $\pm$ 0.40 & 56 $\pm$ 28 & 589 $\pm$ 75
\enddata
\tablecomments{The format follows that of Table 3, except that we show single (FWHM\,$\le$\,10000\kms) component models from regions with line S/N$<$5 (using Ly$\alpha$ for 1150--2000\AA, and \ion{Mg}{2} for 2000--3500\AA) along with the broad component models for objects with S/N$\ge$5. By construction, some single components show narrow (FWHM\,$\le$\,1200\kms) widths.} 
\end{deluxetable*} 
\endgroup

\begingroup
\setlength{\tabcolsep}{3pt}
\begin{deluxetable*}{ccccccccccc}
\tabletypesize{\scriptsize}
\tablecaption{Rest near-UV Emission Line Properties}
\tablehead{
\colhead{Name} & \colhead{$\log L$} & \colhead{$\Delta v$} & \colhead{FWHM} & \colhead{$\sigma$} & \colhead{$\log L_{\rm bl}$} & \colhead{$\Delta v_{\rm bl}$} & \colhead{FWHM$_{\rm bl}$} & \colhead{$\log L_{\rm nl}$} & \colhead{$\Delta v_{\rm nl}$} & \colhead{FWHM$_{\rm nl}$}\\
\colhead{} & \colhead{(\ergs)} & \colhead{(\kms)} & \colhead{(\kms)} & \colhead{(\kms)} & \colhead{(\ergs)} & \colhead{(\kms)} & \colhead{(\kms)} & \colhead{(\ergs)} & \colhead{(\kms)} & \colhead{(\kms)}}
\startdata
\sidehead{\ion{Mg}{2}}
W0126--0529 & 42.33 $\pm$ 0.15 & --416 $\pm$ 50 & 439 $\pm$ 114 & 185 $\pm$ 49 & 42.33 $\pm$ 0.15 & --416 $\pm$ 50 & 439 $\pm$ 114 & ... $\pm$ ... & ... $\pm$ ... & ... $\pm$ ...\\
W2042--3245 & 44.35 $\pm$ 0.16 & 0 $\pm$ 618 & 3865 $\pm$ 1085 & 1641 $\pm$ 461 & 44.35 $\pm$ 0.16 & 0 $\pm$ 618 & 3865 $\pm$ 1085 & ... $\pm$ ... & ... $\pm$ ... & ... $\pm$ ...\\
W2136--1631 & 43.39 $\pm$ 0.09 & --296 $\pm$ 55 & 624 $\pm$ 74 & 540 $\pm$ 49 & 42.84 $\pm$ 0.20 & --1050 $\pm$ 70 & 776 $\pm$ 177 & 43.25 $\pm$ 0.09 & --10 $\pm$ 33 & 608 $\pm$ 80\\
\sidehead{[\ion{Ne}{5}]}
W1103--1826 & 44.13 $\pm$ 0.11 & --1943 $\pm$ 47 & 5853 $\pm$ 1132 & 2485 $\pm$ 481 & 44.13 $\pm$ 0.11 & --1943 $\pm$ 47 & 5853 $\pm$ 1132 &  ... $\pm$ ... &  ... $\pm$ ... &  ... $\pm$ ...\\
W2026+0716 & 43.86 $\pm$ 0.16 & --1179 $\pm$ 160 & 1373 $\pm$ 377 & 583 $\pm$ 160 & 43.86 $\pm$ 0.16 & --1179 $\pm$ 160 & 1373 $\pm$ 377 & ... $\pm$ ... & ... $\pm$ ... & ... $\pm$ ...\\
W2136--1631 & 43.58 $\pm$ 0.07 & --507 $\pm$ 39 & 1535 $\pm$ 1059 & 588 $\pm$ 33 & 43.21 $\pm$ 0.12 & --1050 $\pm$ 69 & 777 $\pm$ 177 & 43.33 $\pm$ 0.07 & --9 $\pm$ 33 & 609 $\pm$ 80
\enddata
\tablecomments{The format follows that of Table 3, with the same exception as Table 4.} 
\end{deluxetable*} 
\endgroup

In Figure 3 we plot the fits to the spectra for well-detected (S/N$\ge$3) emission lines, and list the line properties in Tables 3--5. The detection rates for the strongest lines more than half detected, are 5/9 (Ly$\alpha$), 9/11 ([\ion{O}{2}]), 8/12 (H$\beta$), 11/12 ([\ion{O}{3}]), 8/8 (H$\alpha$). The corresponding numbers for the weaker lines are 3/9 (\ion{N}{5}), 1/9 (\ion{C}{4}), 1/10 (\ion{C}{3}]), 3/10 (\ion{Mg}{2}), 3/10 ([\ion{Ne}{5}]), 0/11 (H$\gamma$), 4/8 ([\ion{O}{1}]), 4/8 ([\ion{N}{2}]), 3/8 ([\ion{S}{2}]). We focus on the stronger lines for the statistical analysis of the ionized gas outflows and broad line kinematics, and use S/N$\ge$5 or $\ge$\,3 cuts depending on the parameter of interest. 

\section{Results and Discussion}
\subsection{Redshift}
The first goal for the XSHOOTER observations was to identify undetermined redshifts, and to confirm the redshifts of some sources with existing but uncertain optical spectroscopy (Eisenhardt et al., in preparation). We define quality ``A'' and ``B'' redshifts as those derived from sources with two or more lines with S/N$\ge$5 and $\ge$\,2, respectively. The systemic redshift is determined by the peak of the model fit around the prominent emission lines (from the common narrow component center of the H$\alpha$, H$\beta$, [\ion{O}{2}] lines, and that of \ion{Mg}{2}, [\ion{Ne}{5}]), which should suffer less from systematic blueshifts (\ion{C}{4}) or blending issues (Ly$\alpha$, \ion{N}{5}, \ion{C}{3}], H$\gamma$). We combined the UV and optical redshifts from S/N$\ge$3 detections and performed error-weighted averaging to calculate the systemic redshifts and uncertainty values. Comparing the systemic (redshift with median uncertainty of 0.00047) to the individual redshifts from the total profile of the stronger lines, we find that the latter lies mostly within three times its uncertainty to the systemic from the [\ion{O}{2}] line with S/N$\ge$3 (8/9), but not from the H$\beta$ (3/8), [\ion{O}{3}] (1/11), H$\alpha$ (3/8) lines where they often show clear blueshifts (\S4.4, \S4.5).

Out of the 10 XSHOOTER spectra we determine two new redshifts (quality ``A'' for W0114--0812, quality ``B'' for W2016--0041), and study 8 with existing values (Eisenhardt et al., in preparation, including two redshifts added in \citealt{Tsa15}). Six of these match within 1\% to existing values (all quality ``A'' redshifts). However, two differ drastically: quality ``A'' $z=0.8301$ for W0126--0529, previously $z=2.937$, and quality ``B'' $z=2.958$ for W2042--3245\footnote{We have S/N=2.0 (Ly$\alpha$), 1.8 (\ion{C}{4}), and 4.2 (\ion{Mg}{2}) from XSHOOTER, but with weak S/N, this is still classified as quality ``B''.}, previously $z=3.963$. The two mismatches occur for redshifts previously determined from Gemini/GMOS observed-frame 4000--6500\AA\ spectra \citep{Tsa15}. Both were quality ``B'' redshifts, derived by the detection, but uncertain identification, of a single feature. The broadband VLT spectra emphasizes the importance of multiple line detections for confident redshift determinations, although our quality ``B'' redshifts could still be biased by noise and we remove those sources from further analysis. As a side note, the two redshifts (W1136+4236, W2216+0723) from \citet{Wu18} were not flagged in their work, but are both quality ``A'' by our definition, showing clear detection of multiple lines.

\begin{figure*}
\centering
\includegraphics[scale=.67]{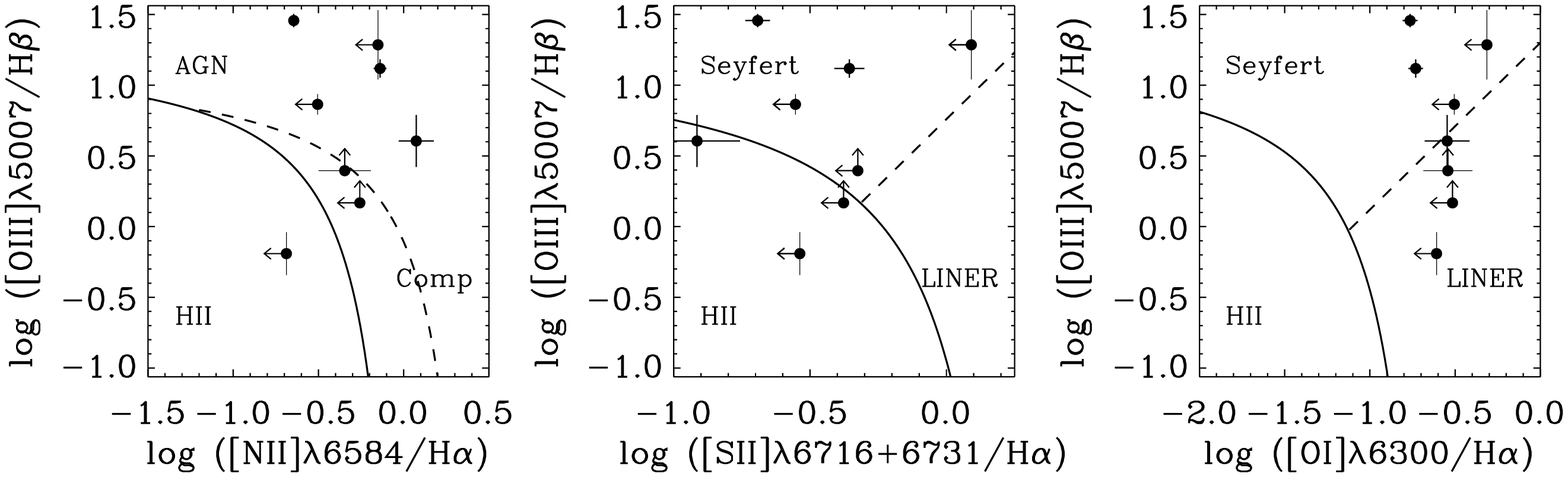}
\caption{BPT diagrams of the sample. Black dots are the line ratios derived from the narrow line fluxes with S/N$\ge$3, and arrows indicate the ratios based on 3-$\sigma$ upper limits. Solid lines denote the dividing lines between star-forming and AGN ratios for all panels \citep{Kew01}, the dashed line between pure AGN and composite star-forming and AGN ratios for the left panel \citep{Kau03}, and between Seyfert and LINER type AGN ratios for the center and right panels \citep{Kew06}. The axes are scaled to follow the observed distribution of typical AGN (e.g., \citealt{Kew06}).}
\end{figure*}

\begin{deluxetable}{ccccc}
\tabletypesize{\scriptsize}
\tablecaption{Seyfert Type}
\tablehead{
\colhead{Name} & \colhead{Seyfert} & \colhead{} & \colhead{} & \colhead{}\\
\colhead{} & \colhead{H$\beta$/[\ion{O}{3}]} & \colhead{broad H$\beta$} & \colhead{broad H$\alpha$} & \colhead{Combined}}
\startdata
W0114--0812 & $\ge$1.8 & no & yes & 1.9\\
W0126--0529 & 1.2--1.5 & no & yes & ...\\
W0147--0923 & $\ge$1.5 & ... & no & 2.0\\
W0226+0514 & $\ge$1.5  & ... & no & 2.0\\
W1103--1826 & 1.5 & ...	& yes & 1.5\\
W1136+4236 & $\ge$1.8 & ... & yes &1.8--1.9\\
W1719+0446 & $\ge$1.2 & ... & ... & $\ge$1.2\\
W2026+0716 & $\ge$1.8 & ... & ... & $\ge$1.8\\
W2136--1631 & $\ge$1.8 & yes	& yes & 1.8\\
W2216+0723 & 1.5 & yes & yes & 1.5	 
\enddata
\tablecomments{The Seyfert types are determined by the H$\beta$/[\ion{O}{3}] ratio (type 1.0, 1.2, 1.5, $\ge$1.8, divided by H$\beta$/[\ion{O}{3}] ratios of 5, 2, 1/3), and the detectability of broad Balmer lines ($\le$1.8 if H$\beta$ and H$\alpha$ are present, 1.9 if only H$\alpha$ is, and 2.0 if both are absent), following \citet{Win92}. Ranges in the values are given taking into account the uncertainty in the H$\beta$/[\ion{O}{3}] ratio, 3-sigma upper limit on H$\beta$ flux ([\ion{O}{3}] instead of H$\beta$ for W1719+0446 due to complete undetection), and the absence of broad Balmer line detectability at S/N$<$5.}
\end{deluxetable} 

\begin{deluxetable*}{ccccccccc}
\tabletypesize{\scriptsize}
\tablecaption{BPT Classification}
\tablehead{
\colhead{Name} &\colhead{BPT ratios} & \colhead{} & \colhead{} & \colhead{} & \colhead{BPT classification}\\
\colhead{} & \colhead{[\ion{O}{3}]/H$\beta$} & \colhead{[\ion{N}{2}]/H$\alpha$} & \colhead{[\ion{S}{2}]/H$\alpha$} & \colhead{[\ion{O}{1}]/H$\alpha$} & \colhead{[\ion{N}{2}]} & \colhead{[\ion{S}{2}]} & \colhead{[\ion{O}{1}]} & \colhead {Combined}}
\startdata
W0114--0812 &  0.86 $\pm$ 0.07 & $<$--0.50 & $<$--0.55 & $<$--0.50 & AGN/\ion{H}{2} & Sy & Sy & Sy\\
W0126--0529 &  --0.19 $\pm$ 0.15 &  $<$--0.69 & $<$--0.54 & $<$--0.61 & \ion{H}{2} & \ion{H}{2} & \ion{H}{2}/LI & \ion{H}{2}\\
W0147--0923 &  $>$0.17 & $<$--0.26 & $<$--0.38 & $<$--0.52 & Any & \ion{H}{2}/Sy & Any & Any\\
W0226+0514 &  $>$0.40 & --0.35 $\pm$ 0.15 & $<$--0.32 & --0.54 $\pm$ 0.14 & Comp/AGN & \ion{H}{2}/Sy	& Sy/LI 			& Sy\\
W1103--1826 &  1.29 $\pm$ 0.24 & $<$--0.15 & $<$0.09 & $<$--0.31 & AGN & Sy & Sy & Sy\\
W1136+4236 &  1.12 $\pm$ 0.06 & --0.14 $\pm$ 0.04 & --0.36 $\pm$ 0.05 & --0.73 $\pm$ 0.04 & AGN	& Sy & Sy & Sy\\
W1719+0446 & $>$--0.33 & ...& ... & ... & ... & ... & ... & ...\\
W2026+0716 & $>$1.56 & ...& ... & ... & AGN & Sy	 & Sy & Sy\\
W2136--1631 & 1.46 $\pm$ 0.05 & --0.64 $\pm$ 0.03 & --0.69 $\pm$ 0.05 & --0.76 $\pm$ 0.04 & AGN & Sy & Sy & Sy\\
W2216+0723 & 0.60 $\pm$ 0.18 & 0.07 $\pm$ 0.10 & --0.91 $\pm$ 0.16 & --0.55 $\pm$ 0.13 & AGN & \ion{H}{2}/Sy & Sy/LI & Sy
\enddata
\tablecomments{The BPT types Composite, Seyfert, and LINER are abbreviated as Comp, Sy, and LI. Empty values correspond to the line not being covered. Upper/lower limits are based on 3-$\sigma$ detection limits unless the line is completely undetected, where we use 3-$\sigma$ limits for H$\alpha$ instead of [\ion{N}{2}] or [\ion{O}{1}] for W0114--0812, [\ion{O}{3}] instead of H$\beta$ for W1719+0446. The combined BPT classification is determined by the majority of the individual classifications.}
\end{deluxetable*} 

\subsection{Spectral Classification}
We can obtain spectroscopic classifications of our targets, such as the Seyfert type\footnote{Throughout, we follow the classic term Seyfert to distinguish from \ion{H}{2} regions or LINERs, but note that the luminosities of Hot DOGs fall into the quasar regime.} (e.g., \citealt{Ost81}), Baldwin, Phillips \& Terlevich (hereafter BPT) diagram (e.g., \citealt{Bal81}), [\ion{O}{2}]-to-[\ion{O}{3}] line ratio (e.g., \citealt{Fer86}), and H$\alpha$-to-H$\beta$ ratio (e.g., \citealt{Ost89}), to better differentiate between the AGN and star-forming properties of Hot DOGs. According to the narrow-to-broad line flux ratios for Balmer lines (assuming the broad lines come from the BLR, not from outflows), all targets with sufficient emission line S/N ($\ge$\,5) in H$\alpha$ (N=8) or H$\beta$ (N=4) show a clear narrow component with no pure type 1 (broad-line dominated) classification. We calculate the Seyfert types using the total [\ion{O}{3}]/H$\beta$ ratio and detectability of broad Balmer lines \citep{Win92}, using S/N$\ge$3 detections (and 3-sigma upper limits) for the [\ion{O}{3}]/H$\beta$ ratio and S/N$\ge$5 detections for the detectability of broad Balmer lines. The Seyfert types shown in Table 6 range from type 1.5 (intermediate H$\beta$ to [\ion{O}{3}], 2/8), to type 1.8 and higher (weak H$\beta$ to [\ion{O}{3}], 6/8), apart from the undetermined type for W0126--0529 (which turns out not to be an AGN, Table 7) and a loosely determined type for W1719+0446 (type $\ge$1.2). Our type 1 fraction (2/8 counting up to type 1.5, but 3--5/8 when including type 1.8--1.9 sources as type 1), together with a majority of single broad Balmer line FWHM values of $\lesssim$3000\kms\ in Table 3, are marginally different from other red quasar spectra in the literature, which show type 1 fractions of $\gtrsim$50--57\% (e.g., \citealt{Gli12}; \citealt{Ban15}; \citealt{Ros15}) with a single broad line FWHM $\gtrsim$3000\kms. This difference likely arises because our targets have higher extinction (\S2.1), and hence are obscuring more of the broad line region. 

Using the narrow line ($\rm FWHM<1200\kms$) ratios with S/N$\ge$3 detections or 3-$\sigma$ upper limits, we plot the [\ion{N}{2}], [\ion{S}{2}], and [\ion{O}{1}] BPT diagrams in Figure 4. There are 8 values in each panel, and two lower limits on [\ion{O}{3}]/H$\beta$ outside the plots based on the absence of [\ion{N}{2}]/H$\alpha$ (W1719+0446, W2026+0716). For $E(B-V)\lesssim1$ (e.g., Table 8) and a Milky way extinction curve, the [\ion{N}{2}]/H$\alpha$, [\ion{S}{2}]/H$\alpha$, [\ion{O}{1}]/H$\alpha$, and [\ion{O}{3}]/H$\beta$ change by less than 0.004, 0.03, $-$0.06, and 0.06 dex if corrected for extinction, which is within the measurement uncertainties. We list each and combined classifications based on the most number of overlaps in Table 7. Combined, seven are BPT AGN, two are unconstrained (W0147--0923, W1719+0446), and the remaining object is star-forming (W0126--0529). The latter source, the lowest redshift galaxy in our sample, is not considered in the following when deducing and discussing AGN properties of our targets. Considering the [\ion{S}{2}]$\lambda$6716, 6731 and [\ion{O}{1}]$\lambda$6300 BPT diagrams, the remaining targets always favor the so-called Seyfert region of these diagrams over the LINER region, implying that Hot DOGs have a strong ionization source from the AGN. 

We independently check for signs of star formation activity in our sample using the [\ion{O}{2}]/[\ion{O}{3}] line ratio (e.g., \citealt{Ho05}; \citealt{Kim06}), as the [\ion{O}{2}] emission usually originates from star formation rather than the AGN activity, and conversely for the [\ion{O}{3}]. The [\ion{O}{2}]/[\ion{O}{3}] ratio values based on S/N$\ge$3 detections are listed in Table 8. Indeed, out of eight objects with both BPT classification and [\ion{O}{2}]/[\ion{O}{3}] ratio values, three are classified exclusively as AGN in all three BPT diagrams have [\ion{O}{2}]/[\ion{O}{3}] values 0.1--0.4, and five with at least one potential \ion{H}{2} BPT classification have [\ion{O}{2}]/[\ion{O}{3}] values 0.2--2.6. This is consistent with pure AGN on the BPT diagrams showing stronger [\ion{O}{3}] to those with potentially mixed star formation with stronger [\ion{O}{2}]. The diverse range of [\ion{O}{2}]/[\ion{O}{3}] ratios are in line with a combination of AGN and star-forming galaxy templates explaining the photometric SEDs of Hot DOGs (e.g., \citealt{Ass15}), but we caution against any conclusive interpretation of the AGN/star-forming nature of the oxygen lines as the [\ion{O}{2}] line could also originate from the AGN activity (e.g., \citealt{Yan06}; \citealt{Mad18}; \S4.5). Overall, the majority (7/8) of Hot DOGs are best explained as non-LINER AGN on the BPT diagrams, and AGN activity dominates star formation in both the BPT and [\ion{O}{2}]/[\ion{O}{3}] diagnostics.

\subsection{Extinction measures}
We next investigate the level of obscuration within the narrow line region using the Balmer decrements (narrow H$\alpha$/H$\beta$) listed in Table 8. Observationally, \citet{Kim06} show that type 1 AGN generally have Balmer decrements close to the expected value, with a distribution sharply peaked at H$\alpha$/H$\beta$=3.3, while \citet{Zak03} show that type 2 AGN have higher values, fwith an average of 4.1 measured from their composite spectrum. Most (7/8) of our sources show Balmer decrements $\gtrsim$4, indicating large amounts of dust obscuration. The translated $E(B-V)$ values assuming intrinsic Balmer decrements of 3.1 expected for AGN narrow line regions (e.g., \citealt{Ost89}), lie mostly around 0.3--0.7 mag, which is not only an order of magnitude higher than the type 1 AGN values (0.06 mag) in \citet{Kim06}, but also up to several times higher than that of the type 2 quasars at various redshifts ($\sim$0.3 mag, \citealt{Zak03}; \citealt{Gre14}). Interestingly, the $E(B-V)$ values from the broad-band SED fitting (3--20 mag, Table 1) are another order of magnitude larger than the values inferred from the Balmer decrements (Table 8). This has been already seen (\citealt{Zak03,Zak05}; \citealt{Gre14}) from similar measurements of candidate type 2 quasars, altogether suggesting a dense, stratified distribution of dust between the compact accretion disk and the extended narrow line region in quasars. This is also consistent with significant amounts of scattered continuum light from unobscured lines of sight in some Hot DOGs or extremely red quasars (e.g., \citealt{Ass16}; \citealt{Ham17}). 

Hot DOGs therefore seem to require not only a dense source of extinction interior to narrow line region, but also an extended distribution of gas or dust outside it. The overall dust temperature of Hot DOGs (50--120K, e.g., \citealt{Wu12}; \citealt{Bri13}) is marginally higher than starburst galaxies, implying that the dust may be associated with the AGN activity. One possibility is to consider host galaxy dust (e.g., \citealt{Rig06}; \citealt{Pol08}). As favored by merger-driven quasar fueling models and observations of Hot DOGs (e.g., \citealt{Hop08}; \citealt{Fan16}; \citealt{Far17}), the obscured starburst galaxy will receive feedback from the quasar activity and become unobscured (see also, \citealt{Buc17}; \citealt{Gob18} for discussion of how the gas content in $z\sim2$ galaxies is higher than in local galaxies). We investigate whether the kpc-scale obscuration is responsible for the reddening by estimating the spatial extent of the narrow line region. We use the narrow line region size--luminosity ($R_{\rm NLR}$--$L_{\rm [O\,III]}$) relation at extinction-uncorrected $L_{\rm [O\,III]}\sim 10^{43.5-45} \ergs$. This reaches the upper limit of $\sim$10\,kpc extrapolated from lower luminosities (e.g., \citealt{Hus13}; \citealt{Hai14}; \citealt{Liu14}; \citealt{Sto18}), comparable to or even larger than the size of the entire host galaxy. Thus, the extinction measured from the narrow line region is more likely coming from the host galaxy, rather than from the central AGN. 

Independent of the Balmer decrement based $E(B-V)$ estimates, we list the extinction-uncorrected, spectroscopically derived [\ion{O}{3}] rest-frame equivalent width (EW) values in Table 8. We compare the EW$_{[\rm O\,III]}$ values for Hot DOGs to type 1 and 2 quasars at matched luminosity ($L_{\rm [O\,III]}\gtrsim 10^{43.5} \ergs$ from \citealt{Gre14}; $L_{5100}\gtrsim 10^{46.5} \ergs$ from \citealt{She16}). It appears there is a large discrepancy in the observed values - the type 1 EW$_{[\rm O\,III]}$ values are $\sim$100\AA\ in \citet{Gre14} 
as opposed to $\sim$10\AA\ in \citet{She16}. However, \citet{She16} note the anti-correlation between the EW$_{[\rm O\,III]}$ and $L_{5100}$ ([\ion{O}{3}] Baldwin effect, e.g., \citealt{Bal77}; \citealt{Bro96}), and as the data points in \citet{Gre14} are from $L_{5100}\lesssim 10^{46} \ergs$ quasars from \citet{She11}, we use the value 10\AA\ in \citet{She16} as a reference. Hot DOGs with a non-\ion{H}{2} BPT classification have EW$_{[\rm O\,III]}$ values around 10--400 \AA\ (median=44 \AA), with corresponding $E(B-V)$ between the line of sights through accretion disk and the [\ion{O}{3}] region of 0.03--1.14 mag (median 0.46 mag), assuming a Milky way extinction curve. Note that we expect the intrinsic EW$_{[\rm O\,III]}$ values (and thus the inferred extinction values) to be higher than measured since these heavily obscured AGN are expected to have non-negligible host galaxy contributions to their continuum emission.

\subsection{Ionized [\ion{O}{3}] gas outflows}
In Figure 5 we plot model profiles of the strongest lines - [\ion{O}{2}], H$\beta$/[\ion{O}{3}], and H$\alpha$, with their broad and narrow components highlighted. It is evident that most of the sample displays broadened or blueshifted [\ion{O}{3}] indicative of ionized gas outflows, often modeled by biconical motions where the blue wing is less affected by obscuration than the red, leading to an asymmetric line profile (e.g., \citealt{Cre10}; \citealt{Zak14}; \citealt{Bae16}). To quantify the presence of outflows as clear non-gravitational motion to our line-of-sight, we first require the presence of a broad [\ion{O}{3}] component\footnote{When we refer to $\Delta v$ or $\sigma$ of a Gaussian component ($\Delta v_{\rm broad/narrow}$, $\sigma_{\rm broad/narrow}$), it is directly converted as the model center and FWHM/2.355 respectively, whereas those of the total profile ($\Delta v_{\rm total}$, $\sigma_{\rm total}$) are calculated as the luminosity weighted 1st or 2nd moments of the model unless specified otherwise.} with $\sigma_{[\rm O\,III], broad}>400\kms,$ to distinguish potential broadening by even the most massive galaxy potential. We also place a S/N$\ge$5 requirement on the [\ion{O}{3}] line to enable robust separation of the broad component from the narrow component (as we did in \S4.2), since line modeling is susceptible to noise at low S/N. The fraction of [\ion{O}{3}] outflows is 8/9. 

\begingroup
\setlength{\tabcolsep}{3pt}
\begin{deluxetable}{ccccc}
\tabletypesize{\scriptsize}
\tablecaption{Line ratios and equivalent widths}
\tablehead{
\colhead{Name}  & \colhead{[\ion{O}{2}]/[\ion{O}{3}]} & \colhead{H$\alpha$/H$\beta$} & \colhead{$E(B-V)$} & \colhead{EW$_{[\rm O\,III]}$}\\
\colhead{} & \colhead{} & \colhead{} & \colhead{(mag)} & \colhead{(\AA)}}
\startdata
W0114--0812 &  0.23 $\pm$ 0.04 & 2.85 $\pm$ 0.51 & $-$0.09 $\pm$ 0.18 & 259.1 $\pm$ 59.8\\
W0126--0529 &  2.62 $\pm$ 0.90 & 4.06 $\pm$ 0.94 & ... & 4.6 $\pm$ 0.6\\
W0147--0923 &  2.47 $\pm$ 0.84 & $>$5.23 & $>$0.53 & 11.1 $\pm$ 2.1\\
W0226+0514 & 1.00 $\pm$ 0.35 & $>$8.32 & $>$1.00 & 25.8 $\pm$ 7.5\\
W1103--1826 & 0.42 $\pm$ 0.25 & 8.12 $\pm$ 4.86 & 0.98 $\pm$ 0.61 & 43.8 $\pm$ 6.7\\
W1136+4236 & ... & 5.79 $\pm$ 0.84 & 0.63 $\pm$ 0.15 & 190.9 $\pm$ 29.2\\
W1719+0446 & 0.64 $\pm$ 0.50 & ... &  ... &  17.5 $\pm$ 2.9\\
W2026+0716 & 0.16 $\pm$ 0.03 & ... & ... & 207.1 $\pm$ 54.6\\
W2136--1631 & 0.08 $\pm$ 0.01 & 4.29 $\pm$ 0.47 & 0.33 $\pm$ 0.11 & 384.5 $\pm$ 32.0\\
W2216+0723 & 1.63 $\pm$ 0.51 & 6.42 $\pm$ 2.78 & 0.74 $\pm$ 0.44 & 36.0 $\pm$ 4.2
\enddata
\tablecomments{Lower limits correspond to 3-$\sigma$ upper limits, and empty values indicate the line not being covered. The $E(B-V)$ values are determined from the measured narrow H$\alpha$/H$\beta$ values assuming an intrinsic Balmer decrement of 3.1, except for W0126--0529 being a BPT non-AGN.}
\end{deluxetable} 
\endgroup 

\begin{figure*}
\centering
\includegraphics[scale=.95]{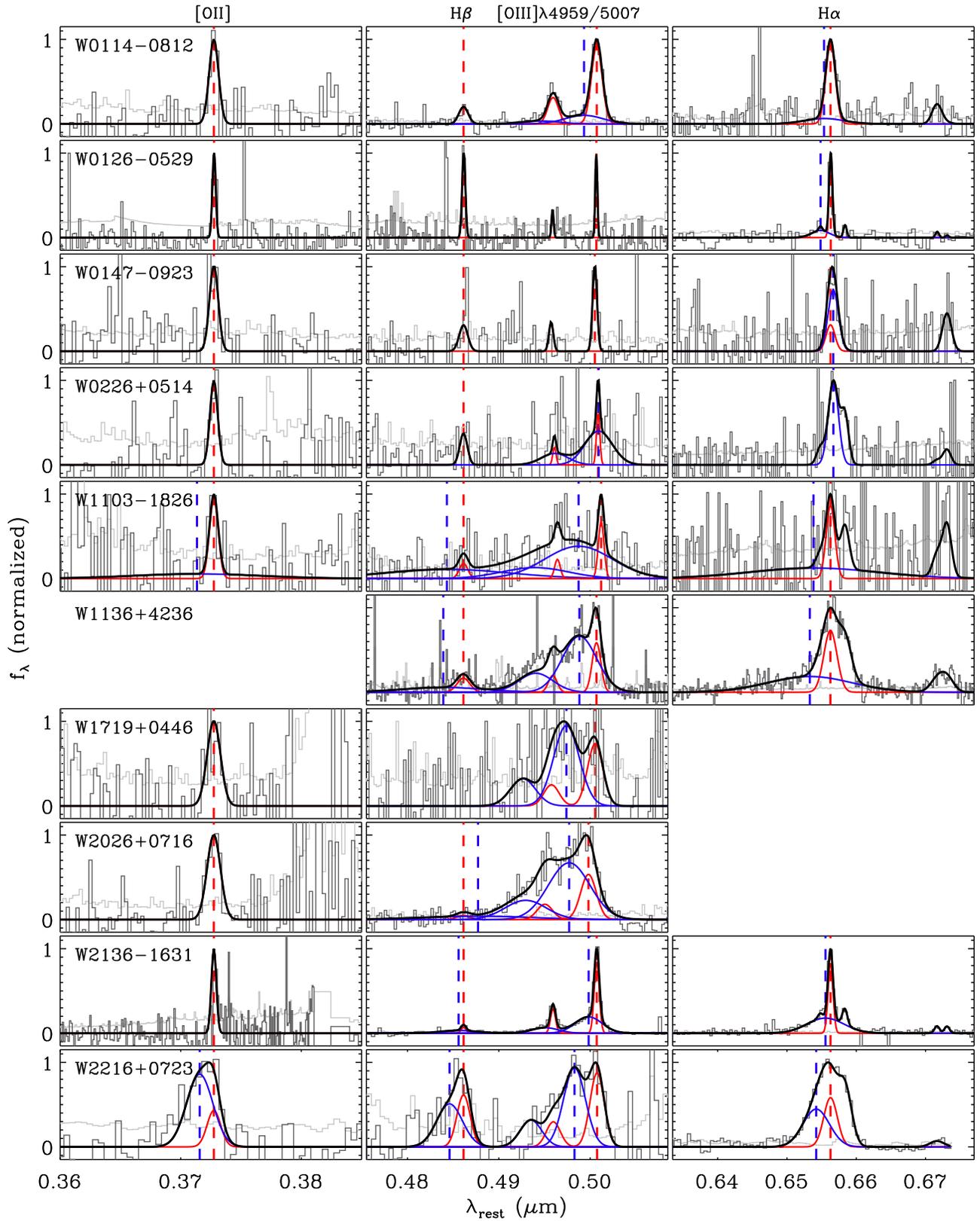}
\caption{Model fit to the [\ion{O}{2}], H$\beta$/[\ion{O}{3}], and H$\alpha$ regions revisited. The format follows that of Figure 3, with additional red and blue vertical dashed lines indicating the center of the narrow and broad model components, respectively. The panels are scaled to show $\pm$\,10000\kms\ from their centers.}
\end{figure*}

It is well known that outflows traced by [\ion{O}{3}] line widths of $\sigma_{[\rm O\,III], broad}>400\kms$ are prevalent in average luminosity, type 1/2 quasars from the SDSS survey (e.g., \citealt{Mul13}; \citealt{Woo16}; \citealt{Rak18}), but we test whether the [\ion{O}{3}] kinematics of Hot DOGs indicate higher $\sigma_{[\rm O\,III], broad}$ values at their high luminosities. In Figure 6 we plot the broad [\ion{O}{3}] line width and shift to the systemic redshift as a function of the observed (extinction-uncorrected) total [\ion{O}{3}] luminosity for our targets, and compare them to other quasars in the literature. Among $L_{[\rm O\,III]}>10^{42}$\ergs\ quasars in the figure, 74--79\% have $\sigma_{[\rm O\,III], broad}>400\kms$, consistent with the majority of local AGN showing outflows irrespective of type at $L_{[\rm O\,III]}>10^{42}$\ergs\ (e.g., \citealt{Woo16}; \citealt{Rak18}). However, the eight Hot DOGs with [\ion{O}{3}] outflows not only have $\sigma_{[\rm O\,III], broad}$\,=700--2200 \kms\ (median 1100 \kms), $\sigma_{[\rm O\,III], total}$\,=600--2000 \kms\ (median 1100 \kms), and line shifts of $\Delta v_{[\rm O\,III], broad}$\,=$-$2000--100 \kms\ (median $-$1100 \kms). These are much larger than the observed limits $\sigma_{[\rm O\,III]}$\,=500--600 \kms\ and the $\Delta v_{[\rm O\,III]}$\,=$-$(500--600) \kms\ irrespective of whether we use the broad or the total [\ion{O}{3}] profile, from $L_{[\rm O\,III]}<10^{43}$\ergs\ AGN in Figure 6, or from the literature (e.g., \citealt{Woo16}; \citealt{Rak18}). 

\begin{figure*}
\centering
\includegraphics[scale=1]{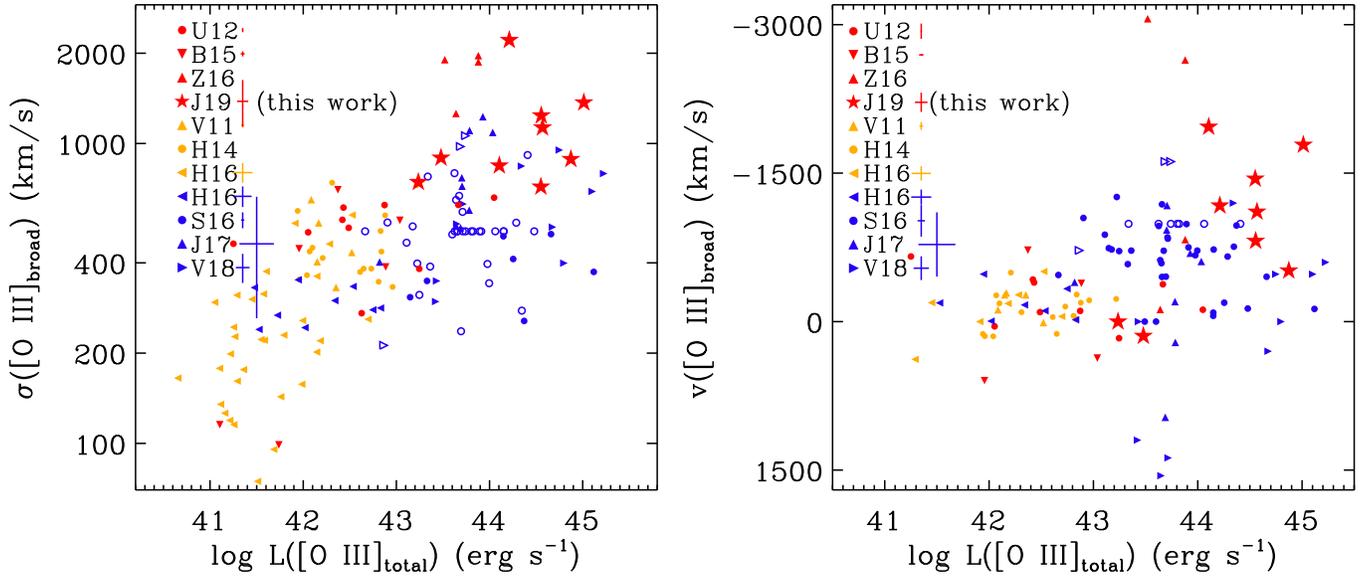}
\caption{The broad [\ion{O}{3}] width ($\sigma$, from FWHM/2.355) and offset against observed (extinction-uncorrected) total [\ion{O}{3}] luminosity. Our sample (red stars) are plotted together with literature data of type 1 (blue), type 2 (yelow), and obscured (red) quasars. We limit the data to those fit by a single narrow and a broad model decomposition of the [\ion{O}{3}], keeping the narrow line width when the broad is absent. Open symbols are those where the best-fit model parameters are on their fitting boundary (S16; V18), or the total (broad+narrow) $\sigma$ values for objects with $\sigma_{\rm total}>\sigma_{\rm broad}$ (S16). When uncertainties are available, only the data with uncertainty in the [\ion{O}{3}] luminosity smaller than 20\% and in the width smaller than 100\%, are plotted. Median uncertainties for each sample are shown when available, with the $\sigma$ uncertainties scaled to when $\sigma_{[\rm O\,III], broad}$\,=800 \kms, as the left panel is a log plot. All luminosities are corrected to our adopted cosmology (\S1).}
\end{figure*}

In Figure 6, we further compare [\ion{O}{3}] outflow kinematics at high luminosity, among heavily obscured (\citealt{Urr12}; \citealt{Bru15}; \citealt{Zak16}, hereafter U12; B15; Z16), type 2 quasars (\citealt{Vil11}, \citealt{Har14}; \citealt{Har16}, hereafter V11; H14; H16), and unobscured quasars (H16; \citealt{She16}, hereafter S16; \citealt{Jun17}, hereafter J17; \citealt{Vie18} including the data from \citealt{Bis17}, hereafter V18). It is noticeable that Hot DOGs together with reddened quasars from Z16 show comparable, or even stronger [\ion{O}{3}] broadening or blueshift than type 1 quasars at the highest luminosity. If a simple, smooth and toroidal geometry of the obscuring structure aligned with the broad line region were to explain most of the obscuration (e.g., \citealt{Ant93}; \citealt{Urr95}), we would expect Hot DOGs, with their high levels of reddening, to appear as type 2 AGN seen very close to edge-on with a high line-of-sight column density. Indeed, most of the measured Seyfert types are close to type 2 ($\gtrsim 1.8$, Table 6), favoring an edge-on orientation of the AGN structure if the obscuration is well explained by geometry. Assuming the biconical outflows are aligned perpendicular to the dusty torus (e.g., \citealt{Fis14}; \citealt{Mar16}), the edge-on geometry implies low line-of-sight velocity of the outflowing material, minimizing the observed Doppler shift and broadening of the (blueshifted/redshifted cone and thus the total) [\ion{O}{3}] profile (e.g., \citealt{Bae16}). The observed [\ion{O}{3}] kinematics in Figure 6 do not follow this expectation however, but are more consistent with intrinsically higher extinction, expected to produce large blueshifts (e.g., \citealt{Bae16}). Thus we require a different source of obscuration beyond the simple torus (e.g., \citealt{Hon13}; \citealt{Asm16} for polar dust geometry, and, e.g., \citealt{Buc17} for extended obscuration within the host galaxy), or a complex velocity structure within the outflowing [\ion{O}{3}] region (e.g., outflows with a large bicone opening angle or the spherical outflows seen in [\ion{C}{2}] observations of the Hot DOG W2246--0526 from \citealt{Dia16}). Although there is a large scatter in the data, we find that the outflow kinematics in Figure 6 show stronger broadening and blueshift as a function of luminosity or reddening, whereas the differences between type 1 and 2 AGN at a given luminosity are relatively minor. This suggests that the obscured quasar phase is related to the production of strong ionized gas outflows, irrespective of possible inclination effects. 

Using the outflow kinematics of the broad [\ion{O}{3}] profile under the assumption of a uniform, filled spherical/biconical outflow geometry (e.g., \citealt{Mai12}), we estimate the outflow quantities for Hot DOGs -- mass outflow rate ($\dot{M}_{\rm out}$), energy injection rate ($\dot{E}_{\rm out}$), and momentum flux ($\dot{P}_{\rm out}$) -- as follows
\begin{eqnarray}\begin{aligned}
&\dot{M}_{\rm out}=\frac{3M_{\rm gas} v_{\rm out}}{R_{\rm out}}\\
&\dot{E}_{\rm out}=\frac{1}{2}\dot{M}_{\rm out} v_{\rm out}^{2}=\frac{3M_{\rm gas} v_{\rm out}^{3}}{2R_{\rm out}}\\
&\dot{P}_{\rm out}=\dot{M}_{\rm out} v_{\rm out}=\frac{3M_{\rm gas} v_{\rm out}^{2}}{R_{\rm out}}.
\end{aligned}\end{eqnarray}
Ionized gas mass ($M_{\rm gas}$), outflow size ($R_{\rm out}$), and extinction/projection-corrected outflow velocity ($v_{\rm out}$) are derived as follows
\begin{equation}\begin{aligned}
&M_{\rm gas}=4.0 \times 10^{7} M_{\odot} \times\\
&\Big(\frac{C}{10^{[O/H]}}\Big)\Big(\frac{L_{\rm [O\,III], broad}}{10^{44}\,\ergs}\Big) \Big(\frac{\langle n_{\rm e}\rangle}{10^{3}\,\rm cm^{-3}}\Big)^{-1}\\
&R_{\rm out}=R_{\rm out}(L_{\rm [O\,III]}>10^{43}\,\ergs)=3 \,\rm kpc\\
&v_{\rm out}=2\sigma_{0}=2\sqrt{\sigma_{\rm [O\,III], broad}^{2}+v_{\rm [O\,III], \rm broad}^{2}},
\end{aligned}\end{equation}
where $C=\langle n_{\rm e}\rangle^{2}/\langle n_{\rm e}^{2}\rangle$, $n_{\rm e}$ is the electron density, and [O/H] is the metallicity of the gas in solar units.

For Equation (2), we adopt the $M_{\rm gas}$ equations from \citet{Nes11} and \citet{Car15}, and assume $C/10^{[O/H]}=1$, where the equation from both works become identical. As for $R_{\rm out}$, the $R_{\rm NLR}$--$L_{[\rm O\,III]}$ relation saturates at around 10\,kpc beyond $L_{[\rm O\,III]}\gtrsim10^{43}$\ergs\ as noted in \S4.3, but the radius where the outflows are effective can be smaller than the maximal extent of the outflowing [\ion{O}{3}] line emitting region. Observations of luminous quasars at $L_{[\rm O\,III]}\gtrsim10^{43}$\ergs\ range between $\sim$\,1--10\,kpc, and differ upon using the spatial offset ($\sim$1\,kpc, e.g., \citealt{Car15}), or the spatial extent ($\sim$1\,kpc when flux weighted, e.g., \citealt{Kan18}; $\sim$5--10\,kpc when measured kinematically or above a S/N threshold, e.g., \citealt{Can12}; \citealt{Cre15}; \citealt{Per15}; \citealt{Kan18}) of the broad/blueshifted [\ion{O}{3}] component. We use a representative value, 3\,kpc, considering the diversity in the measurement methods, and we are unable to spatially resolve the outflow size with our observations. For the objects on which both of the [\ion{S}{2}] doublet lines are detected with S/N$\ge$3, namely W1136+4236 and W2136--1631, we measure the [\ion{S}{2}]$\lambda6716$/[\ion{S}{2}]$\lambda6731$ ratios of $0.82 \pm 0.19$ and $1.02 \pm 0.21$, respectively. Assuming a 10,000K temperature, the line ratios correspond to electron densities $n_{e}\sim1100$ and $\sim600\,\rm cm^{-3}$, respectively \citep{Ost89}. These values are on the higher end but within the range of measurements for various AGN (100--1000 $\rm cm^{-3}$, e.g., \citealt{Hol06}; \citealt{Nes06}; \citealt{Per15}; \citealt{Kar16}; \citealt{Rak17}, but see also, \citealt{Bar19}). We fix $\langle n_{e} \rangle=300\,\rm cm^{-3}$ in between the boundary of reported values due to the limited number of our measurements, but this may underestimate the outflow quantities by a factor of 2--4 if Hot DOGs turn out to have $n_{e}\sim600-1000\,\rm cm^{-3}$. Lastly, $v_{\rm out}$ is adopted from \citet{Bae16} and \citet{Bae17} as a combination of the velocity dispersion and the offset, to correct for dust extinction and projection effects.

\begin{deluxetable*}{ccccccccc}
\tabletypesize{\scriptsize}
\tablecaption{Outflow quantities}
\tablehead{
\colhead{Name} & \colhead{$\log M_{\rm gas}$} & \colhead{$v_{\rm out}$} & \colhead{$\dot{M}_{\rm out}$} & \colhead{$\log \dot{E}_{\rm out}$} & \colhead{$\log \dot{P}_{\rm out}$} & \colhead{$\dot{M}_{\rm out}/\dot{M}_{\rm acc}$} & \colhead{$\dot{E}_{\rm out}/L_{\rm bol}$} & \colhead{$\dot{P}_{\rm out}c/L_{\rm bol}$}\\
\colhead{} & \colhead{($M_{\odot}$)} & \colhead{(\kms)} & \colhead{($M_{\odot}\,\rm yr^{-1}$)} & \colhead{(\ergs)} & \colhead{(dyn)} & \colhead{} & \colhead{(\%)} & \colhead{}}
\startdata
W0114-0812 & 8.12 $\pm$ 0.12 & 2967 $\pm$ 331 & 404 $\pm$ 119 & 45.05 $\pm$ 0.19 & 36.88 $\pm$ 0.15 & 1.82 $\pm$ 0.55 & 0.09 $\pm$ 0.04 & 0.18 $\pm$ 0.07\\
W0226+0514 & 7.53 $\pm$ 0.13 & 1816 $\pm$ 482 & 63 $\pm$ 25 & 43.81 $\pm$ 0.37 & 35.86 $\pm$ 0.26 & 3.37 $\pm$ 1.76 & 0.06 $\pm$ 0.06 & 0.20 $\pm$ 0.14\\
W1103--1826 & 8.28 $\pm$ 0.09 & 5008 $\pm$ 556 & 968 $\pm$ 231 & 45.88 $\pm$ 0.17 & 37.48 $\pm$ 0.13 & 9.61 $\pm$ 2.65 & 1.34 $\pm$ 0.56 & 1.61 $\pm$ 0.54\\
W1136+4236 & 8.61 $\pm$ 0.03 & 3169 $\pm$ 98 & 1306 $\pm$ 105 & 45.62 $\pm$ 0.05 & 37.42 $\pm$ 0.04 & 5.71 $\pm$ 0.51 & 0.32 $\pm$ 0.04 & 0.60 $\pm$ 0.06\\
W1719+0446 & 8.06 $\pm$ 0.45 & 4283 $\pm$ 225 & 507 $\pm$ 521 & 45.47 $\pm$ 0.45 & 37.14 $\pm$ 0.45 & 6.44 $\pm$ 6.78 & 0.66 $\pm$ 0.70 & 0.92 $\pm$ 0.97\\
W2026+0716 & 9.02 $\pm$ 0.04 & 4502 $\pm$ 149 & 4864 $\pm$ 465 & 46.49 $\pm$ 0.06 & 38.14 $\pm$ 0.05 & 7.65 $\pm$ 0.80 & 0.86 $\pm$ 0.12 & 1.15 $\pm$ 0.14\\
W2136--1631 & 8.74 $\pm$ 0.01 & 2051 $\pm$ 32 & 1143 $\pm$ 39 & 45.18 $\pm$ 0.02 & 37.17 $\pm$ 0.02 & 2.48 $\pm$ 0.09 & 0.06 $\pm$ 0.00 & 0.17 $\pm$ 0.01\\
W2216+0723 & 8.47 $\pm$ 0.10 & 3223 $\pm$ 322 & 971 $\pm$ 242 & 45.50 $\pm$ 0.16 & 37.29 $\pm$ 0.13 & 4.41 $\pm$ 1.33 & 0.26 $\pm$ 0.11 & 0.47 $\pm$ 0.17
\enddata
\tablecomments{Outflow quantities are shown for [\ion{O}{3}] lines with S/N$\ge$5, having a broad component ($\sigma_{[\rm O\,III], broad}>400\kms$). $M_{\rm gas}$ is the ionized, outflowing gas mass estimated using the broad [\ion{O}{3}] profile without extinction correction on $L_{[\rm O\,III], broad}$, $v_{\rm out}$ is the outflow velocity with extinction/projection-correction, $\dot{M}_{\rm out}$ is the mass outflow rate, $\dot{E}_{\rm out}$ is the energy injection rate, $\dot{P}_{\rm out}$ is the momentum flux, $\dot{M}_{\rm acc}$ is the mass accretion rate (calculated as $L_{\rm bol}/\eta c^{2}$ with radiative efficiency $\eta=0.1$), $L_{\rm bol}$ is the bolometric luminosity using a [\ion{O}{3}] to bolometric correction factor of 3500 \citep{Hec04}.} 
\end{deluxetable*} 

There are several systematic uncertainties in using Equations (1)--(2) to derive [\ion{O}{3}] outflow quantities (e.g., \citealt{Har18}). First, we assume the narrow [\ion{O}{3}] comes from a non-outflowing region, and use only the broad [\ion{O}{3}] to measure outflow quantities. This assumption appears valid for our sample as most of the narrow Gaussian components have offsets within $\sim$100\kms\ to the systemic redshift (with exceptions for W1103--1826 and W2026+0716), for non-\ion{H}{2} BPT sources with line S/N$\ge$5 (Table 3 and Figure 5). There are counter examples of narrow emission lines altogether drifting against the stellar absorption line-based redshift for local AGN (e.g., \citealt{Bae16}), but the occurrence is rare, supporting the fact that the narrow component of the [\ion{O}{3}] is less likely to be outflowing. Still, there are 3/9 BPT non-\ion{H}{2} sources with [\ion{O}{3}] S/N$\ge$5 and narrow component widths $\sigma_{[\rm O\,III],narrow}>400\kms$. This is partly due to our narrow line FWHM limit allowing up to 1200\kms\ which is an arbitrary division between a broad and a narrow line (\S3), and we check how much the outflow quantities change if we substitute the broad [\ion{O}{3}] component properties in Equation (2) into those of the total. We find that $M_{\rm gas}$ changes by a factor of 1.30--1.61 and $v_{\rm out}$ changes by a factor of 0.81--0.89. $\dot{M}_{\rm out}$, $\dot{E}_{\rm out}$, $\dot{P}_{\rm out}$ change by a factor of 1.16--1.30, 0.84--0.92, 1.02--1.05, respectively. The overall changes are relatively insensitive to the choice of the broad or the total [\ion{O}{3}] profile, as the change in $M_{\rm gas}$ cancels out with that of $v_{\rm out}$.

Second, we are using measured, i.e., extinction-uncorrected [\ion{O}{3}] luminosities to derive the outflow quantities. Though we take this effect into consideration for the outflow velocity in Equation (2), the stratified distribution of obscuring material between the lines of sight through the AGN center and the narrow line region (Tables 1 and 8) complicates the extinction correction. As the narrow line region is more extended than the region responsible for the majority of the dust obscuration (see \S4.3) we adopt the Balmer decrement-based $E(B-V)$ values to estimate the extinction correction for $L_{\rm [O\,III], broad}$, assuming a similar size of the outflowing [\ion{O}{3}] line region as the narrow Balmer line region. Hot DOGs typically show $E(B-V)\sim$\,0.3--0.7 mag (Table 8), corresponding to the [\ion{O}{3}] extinctions by factors of $\sim$3--9 for the Milky way extinction curve. This reduces the $M_{\rm gas}$ by the same amount. As the objects with Balmer decrement values are limited, we use $L_{[\rm O\,III], broad}$ observed values when deriving $M_{\rm gas}$ and the outflow quantities dependent on $M_{\rm gas}$, but interpret the values considering they are likely underestimated. The outflow quantities normalized by the same extinction-uncorrected luminosity, or the outflow efficiencies, are more reliable under the effect of extinction.

Third, we saw an order of magnitude range in the measurement of both the outflow size and the electron density due to the distribution of measurements and dependence on the measurement method. Assuming our Hot DOGs span the full range of distribution in $R_{\rm out}$ and $\langle n_{\rm e} \rangle$ at our probed luminosities, the uncertainty in the average value of $R_{\rm out}$ and $\langle n_{\rm e} \rangle$ dividing the range by the square root of the number of objects, i.e., 10/$\sqrt{8}$, will be a few times respectively. We thus estimate the uncertainty in the average outflow quantity (proportional to 1/$R_{\rm out} \langle n_{\rm e} \rangle$) to be an order of a magnitude. 

We list the estimated outflow quantities for the eight Hot DOGs with [\ion{O}{3}] S/N$\ge$5 and $\sigma_{[\rm O\,III], broad}>400\kms$ in Table 9. Extinction-uncorrected mass outflow rates are 60--4860 $M_{\odot}\,\rm yr^{-1}$ (median 970 $M_{\odot}\,\rm yr^{-1}$). The estimated star formation rates (hereafter SFRs) for a sample of Hot DOGs based on SED fitting are $\lesssim$300--600 $M_{\odot}\,\rm yr^{-1}$ (e.g., \citealt{Eis12}; \citealt{Jon14}; \citealt{Dia18}), comparable to the starbursts in SMGs at similar redshifts, which show values between 100--1000 $M_{\odot}\,\rm yr^{-1}$ (e.g., \citealt{Mag12}). This implies that Hot DOGs have comparably higher mass outflow rates than SFRs, even when uncorrected for extinction. 

We also obtain the mass loading factor between the mass outflow and accretion, i.e., $\dot{M}_{\rm out}/\dot{M}_{\rm acc}$, with mass accretion rate $\dot{M}_{\rm acc}=L_{\rm bol}/\eta c^{2}$ and radiative efficiency $\eta=0.1$ (e.g., \citealt{Sol82}). Using the [\ion{O}{3}] to bolometric correction factor \citep{Hec04} removes the extinction dependence on $L_{\rm [O\,III]}$ in $\dot{M}_{\rm out}$ when dividing by $L_{\rm bol}$. The $\dot{M}_{\rm out}/\dot{M}_{\rm acc}$ values are 1.8--9.6 (median 5.7) so that the amount of gas outflowing is larger than that accreted, assuming \citet{Hec04} values are appropriate for Hot DOGs.\footnote{The \citet{Hec04} bolometric correction may not direct apply for Hot DOGs if they are a different population than typical AGN. The median and scatter of $L_{\rm bol}$ ratios using bolometric corrections from $L_{\rm [O\,III]}$ and $L_{\rm 5100}$ for eight Hot DOGs with a non-\ion{H}{2} BPT classification, are 0.08$\pm$0.87 dex.} Combined, we find the mass outflow rate of Hot DOGs, with an uncertainty of an order of magnitude or more on the average value, to be marginally greater than the gas consumption due to star formation or fueling the AGN itself, demonstrating the role of ionized gas outflows in depleting the ISM around the AGN, and competing against gas cooling to form additional stars.

\begin{figure}
\centering
\includegraphics[scale=1]{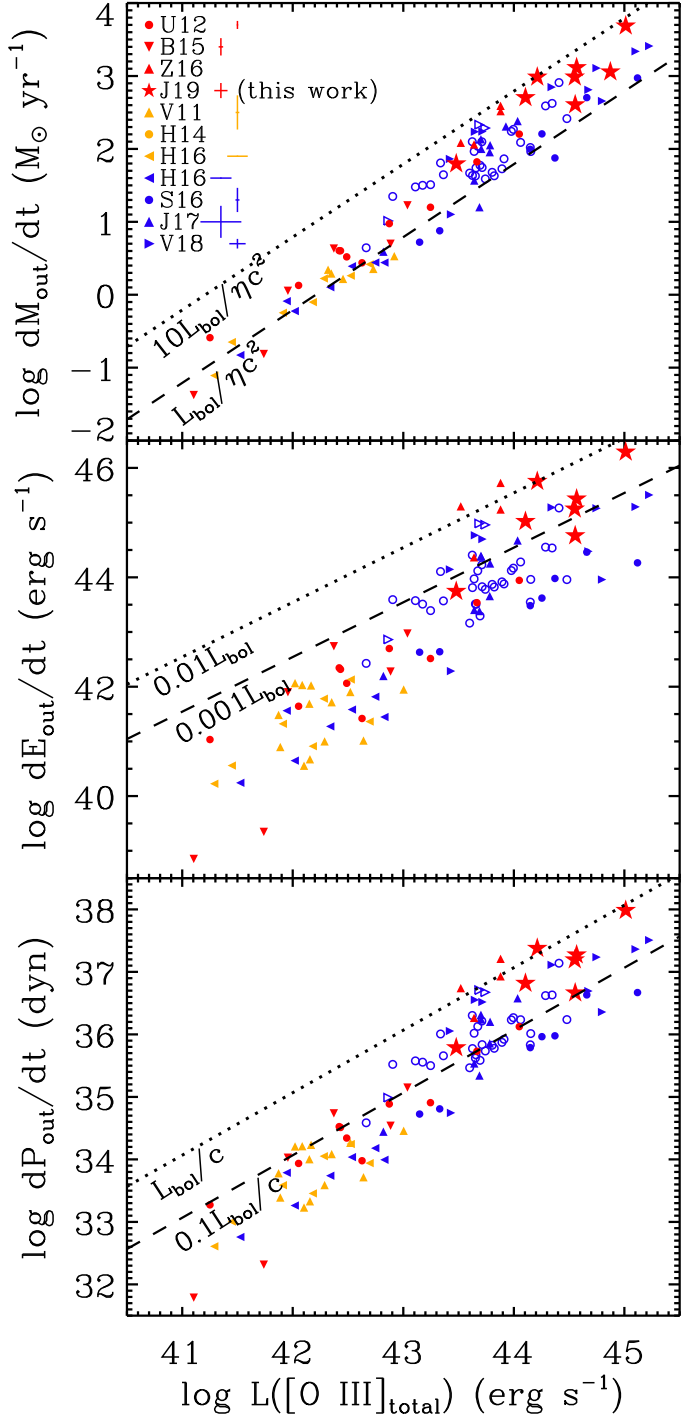}
\caption{The estimated mass outflow rates ($\dot{M}_{\rm out}$), energy injection rates ($\dot{E}_{\rm out}$), and momentum fluxes ($\dot{P}_{\rm out}$) against observed (extinction-uncorrected) total [\ion{O}{3}] luminosity. The colors and symbols follow those of Figure 6, and we show the outflow quantities in constant units of $L_{\rm bol}/\eta c^{2}$, $L_{\rm bol}$, and $L_{\rm bol}/c$, respectively (dotted and dashed lines). Radiative efficiencies of $\eta=0.1$, and [\ion{O}{3}] to bolometric correction factor of 3500 \citep{Hec04} are adopted. When estimating the outflow quantities, we use $\log R_{\rm out} {\rm (kpc)}=0.41\log L_{\rm [O\,III]} (\ergs)-14.00$ from \citet{Bae17} for $\log L_{\rm [O\,III]}<42.63\ergs$, and a fixed $R_{\rm out}$=3 kpc for $\log L_{\rm [O\,III]}\ge42.63\ergs$. This is to match the $R_{\rm out}$ values adopted for Hot DOGs from Equation (2), where $R_{\rm out}(L_{\rm [O\,III]})$ using \citet{Bae17} would exceed 3 kpc.}
\end{figure}

Energy injection rates range from $\dot{E}_{\rm out}=6.5\times10^{43}$--$3.1\times10^{46}$ \ergs\ (median $3.2\times 10^{45}$ \ergs) without extinction correction, also being 0.058--1.3 (median 0.32) \% of $L_{\rm bol}$, independent of the extinction effect. The fraction of the bolometric luminosity transferred to kinetic energy, or the feedback efficiency, are consistent with observations of luminous, obscured quasars (e.g., \citealt{Zak16}), theoretical models and simulations of $\sim10^{3}$\kms\ winds including kinetic energy (e.g., \citealt{Kin03}; \citealt{Cho12}; \citealt{Rot12}), or a weak outflow induced cloud expansion (\citealt{Hop10}). Also, the bolometric luminosity and the fraction of it injected into the ISM are near the limit where it can blowout the gas (e.g., see discussion in \citealt{Dia16}). Our feedback efficiencies are an order or two magnitudes smaller than the thermal efficiencies required to match the normalization of the local BH mass--stellar velocity dispersion ($M_{\rm BH}$--$\sigma_{*}$) relation (5\%, \citealt{Dim05}), or those used in prescriptions for hydrodynamic cosmological simulations ($>$10\%, e.g., \citealt{Dub14}; \citealt{Sch15}; \citealt{Wei17}), but these works have a thermal feedback mode alone, or involve subgrid prescriptions in the simulations due to limited resolution. Momentum flux values are in the range $\dot{P}_{\rm out}=7.2\times10^{35}$--$1.4\times10^{38}$ dyn (median $2.0\times 10^{37}$) dyn, being 0.17--1.6 (median 0.60) times $L_{\rm bol}/c$. Outflows are classified as energy-driven if they retain their thermal energy, and momentum-driven if they radiate away. Following \citet{Tom15} and \citet{Fer15} for energy-conserving outflows observed in local ULIRGs (IRAS F11119+3257 and Mrk 231), the sub-$L_{\rm bol}/c$ momentum flux values for Hot DOGs are comparable to or lower than those for ULIRGs at $v_{\rm out}\sim10^{3}$\kms, indicating that the ionized outflows in Hot DOGs are marginally more likely momentum-driven than energy-driven. We summarize our findings in Figure 7. Hot DOGs show among the strongest outflow quantities ($\dot{M}_{\rm out}$, $\dot{E}_{\rm out}$, $\dot{P}_{\rm out}$) and outflow efficiencies ($\dot{M}_{\rm out}/\dot{M}_{\rm acc}$, $\dot{E}_{\rm out}/L_{\rm bol}$, $\dot{P}_{\rm out}c/L_{\rm bol}$) compared to the AGN samples in the literature.
\begin{figure*}
\centering
\includegraphics[scale=1]{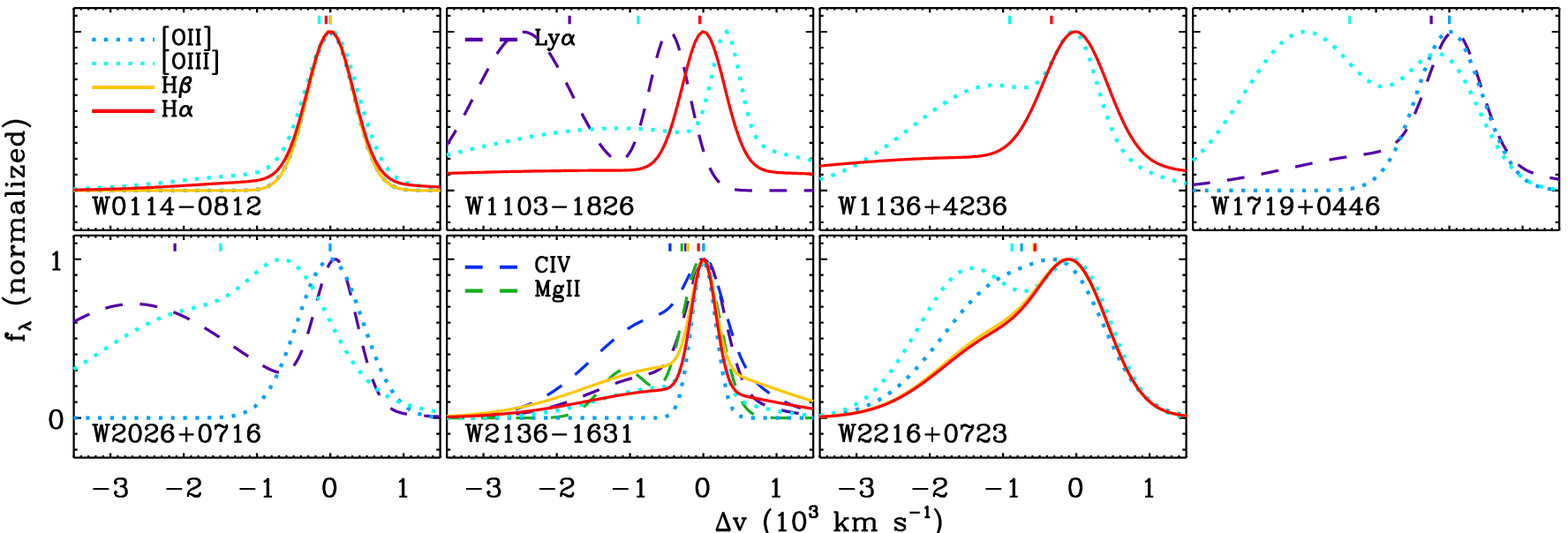}
\caption{The model profiles to the emission lines with S/N$\ge$5 that have either a significantly broad ($\sigma_{\rm broad/narrow}>400\kms$) Balmer line (W0114--0812, W1103--1826, W1136+4236, W2136--1631, W2216+0723) or a [\ion{O}{2}] line (W1719+0446, W2026+0716, W2216+0723). The colors for the lines follow that of Figure 2, and the rest-frame UV lines (Ly$\alpha$-purple, \ion{C}{4}-blue, \ion{Mg}{2}-green) are shown dashed, oxygen lines ([\ion{O}{2}]$\lambda$3727-light blue, [\ion{O}{3}]$\lambda$5007-cyan) dotted, and Balmer lines (H$\beta$-yellow, H$\alpha$-red) solid. The first moment of the total profile is shown on top of the models with corresponding colors.}
\end{figure*}

\subsection{Interpretation of broad lines in obscured AGN}
Having seen the occurrence and strength of [\ion{O}{3}] outflows in our luminous, obscured AGN sample, we now assess the usefulness of broad emission lines in measuring $M_{\rm BH}$ for highly obscured AGN. We overplot in Figure 8 the normalized profiles of prominent lines for the sources with potential outflows in addition to the [\ion{O}{3}] line. The figure includes the 5/7 BPT non-\ion{H}{2} Hot DOGs with significantly broad H$\alpha$ ($\sigma_{\rm H\alpha, broad/narrow}>400\kms$) with S/N$>$5 from Table 3 (W0114--0812, W1103--1826, W1136+4236, W2136--1631, W2216+0723). The normalized broad Balmer line components are weaker than the broad [\ion{O}{3}] but are blueshifted for all five objects, with absolute values comparable to or smaller than the broad [\ion{O}{3}] blueshift with a fraction of 0.53--1.23 times (median 0.65). This trend is also seen in the 1st moment of the Balmer lines ($\Delta v$ in Table 3) being blueshifted, but not as much as that of the [\ion{O}{3}]. 

The broadening of the Balmer line might come from the broad line region, with blueshifts indicating outflows within the broad line region (e.g., \citealt{Vie18}) or the broad line region simply being highly obscured. We check this by calculating the Balmer decrement of objects having S/N$\ge5$ in H$\alpha$ and showing a broad component in both H$\alpha$/H$\beta$ (W1103--1826, W1136+4236, W2136--1631, W2216+0723). The values are 1.93$\pm$1.02, 4.28$\pm$0.87, 1.93$\pm$0.22, and 5.95$\pm$2.17 respectively, all comparable to or smaller than the narrow Balmer decrements in Table 8. This is inconsistent with the expected higher extinction (and thereby higher Balmer decrements) towards the broad line region compared to the narrow line region, assuming that the intrinsic Balmer decrements of the broad line region for Hot DOGs are comparable to the observed values for luminous type 1 quasars at $z\sim2$ (median and scatter of 3.2$\pm$1.4, \citealt{She12}), which themselves are comparable to the intrinsic values for the narrow lines (3.1, e.g., \citealt{Ost89}). Therefore, we find it more likely that broad and blueshifted Balmer lines originate outside the broad line region, and more likely due to outflows within the narrow line region. This is also consistent with 2/2 of the objects not showing a broad H$\alpha$ (2/7) still having $\sigma_{\rm H\alpha, broad/narrow}>300\kms$, which is on the broader side to be explained by non-outflowing motion unless the host galaxies of Hot DOGs have grown to sufficiently massive bulges at $z\sim2$. 

Following \citet{Wu18}, we check if the Balmer line outflows are as strong as the [\ion{O}{3}] outflows by fitting the rest-frame 3500--7000\AA\ spectra with a common broad line center and width, and calculating the F-distribution probability of the reduced chi-square from the fit having a statistically significant improvement. Out of the S/N$\ge5$ spectra, 1/9 (W0147--0923) has an improvement (95\% or higher F-distribution probability) such that the broad Balmer lines are indistinguishable from the outflowing [\ion{O}{3}] component, but 7/9 (including all the broad H$\alpha$ sources in Figure 8) yield significantly improved fits when the broad [\ion{O}{3}] center and width are detached from the rest of the lines. We thus can rule out the case where all the broad Balmer lines are showing outflows as strong as the [\ion{O}{3}], in accord with \citet{Wu18} and consistent with weaker but similar Balmer line kinematics as the [\ion{O}{3}] found in SDSS type 2 quasars (e.g., \citealt{Kan17}). 

In addition to the likelihood of outflows in the narrow Balmer line region for some Hot DOGs, we plot in the remaining panels of Figure 8 the 3/6 BPT non-\ion{H}{2} sources with a significantly broad [\ion{O}{2}] ($\sigma_{\rm [O\,II], broad/narrow}>400\kms$, W1719+0446, W2026+0716, W2216+0723) with S/N$>$5. Additionally, 1/3 of the remaining (3/6) sources (W0114--0812) shows $\sigma_{\rm [O\,II], broad/narrow}>300\kms$. This can either be explained by strong AGN outflows broadening the [\ion{O}{2}] line profile, or the kinematics of the ISM in the star-forming region being highly disturbed by AGN outflows and/or galaxy merging (e.g., \citealt{Dia16,Dia18}). As [\ion{O}{2}] is a forbidden line, a fraction of them being broad adds support for the presence of outflows, and hints for $\sim$kpc scale, narrow line region origin for at least some the broadened Balmer lines.

The presence of outflowing material within the Balmer or even the [\ion{O}{2}] lines can cause serious misinterpretation when using the broad emission line width to measure $M_{\rm BH}$. Although we are limited to a single object (W2136--1631) in Figure 8 covering all the major UV/optical broad emission lines, the Hydrogen lines (Ly$\alpha$, H$\beta$, H$\alpha$) and \ion{Mg}{2} profiles are indistinguishable from [\ion{O}{3}], with \ion{C}{4} showing a stronger broad/blueshifted wing component. This suggests that not only the Balmer lines, but the rest-frame UV broad emission lines could be contaminated by outflows within the narrow line region. Therefore, not only the $M_{\rm BH}$ measurements reported for Hot DOGs (e.g., \citealt{Ric17a}; \citealt{Tsa18}; \citealt{Wu18}), but any $M_{\rm BH}$ measurement for highly obscured AGN with a broad and blueshifted permitted emission line should be carefully tested for the presence of outflows through signatures of broad and blueshifted forbidden lines (e.g., \citealt{Ale08}; \citealt{Ric17c}), especially at $L_{[\rm O\,III]}>10^{42}$\ergs\ (Figure 6). On the other hand, we expect the $M_{\rm BH}$ values for Hot DOGs from emission lines likely coming from scattered light (e.g., \citealt{Ass16,Ass19}) are unlikely affected by outflows.

\subsection{Eddington ratio for Hot DOGs}
We alternatively estimate $M_{\rm BH}$ and the Eddington ratios ($f_{\rm Edd}$) shown in Table 10 using the $M_{\rm BH}$--$\sigma_{*}$ relation and the width of the common narrow component of [\ion{O}{2}]/Balmer lines, noting that $\sigma_{*}$ is similar to narrow emission line widths but with a large uncertainty (a factor of $\sim$2, e.g., \citealt{Gre05a}; \citealt{Zak14}; \citealt{Ben18}). Here, we select the narrow line width from [\ion{O}{2}]/Balmer lines instead of [\ion{O}{3}] to minimize blending from outflows. Considering the dispersion of the $M_{\rm BH}$--$\sigma_{*}$ relation itself (a factor of 2--3; e.g., \citealt{Kor13}; \citealt{Woo13}), and the possibility of a systematic offset for various types of AGN compared to the local inactive galaxies on the BH--galaxy scaling relation (up to a factor of several, e.g., \citealt{Kim08}; \citealt{Urr12}; \citealt{Sex19}), the $M_{\rm BH}$ estimates for Hot DOGs could be inaccurate by up to an order of magnitude. Having in mind those uncertainties, we find most of the $M_{\rm BH}$ values to be in the order of $10^{9} M_{\odot}$ while those associated with $\sigma_{\rm narrow}>400\kms$ are given with upper limits and show values around $\sim10^{10} M_{\odot}$. We convert the $M_{\rm BH}$ estimates and use the $L_{\rm bol}$ from the extinction-corrected 5100\AA\ luminosity with a bolometric correction (e.g., Figure 1) into the Eddington ratio $f_{\rm Edd}=L_{\rm bol}/L_{\rm Edd}$, which ranges in the order of unity to ten. This can be interpreted as the AGN accretion being highly effective near or even beyond its Eddington limit, but there are possibilities of a systematic offset in $f_{\rm Edd}$ for Hot DOGs. 

\begingroup
\setlength{\tabcolsep}{3pt}
\begin{deluxetable}{cccc}
\tabletypesize{\scriptsize}
\tablecaption{Accretion rates}
\tablehead{
\colhead{Name} & \colhead{$\log M_{\rm BH}$} & \colhead{$f_{\rm Edd}$} & \colhead{$f_{\rm Edd} (M_{\rm BH}<10^{10}M_{\odot})$}\\
\colhead{} & \colhead{($M_{\odot}$)} & \colhead{} & \colhead{}}
\startdata
W0114--0812 & 9.34 $\pm$ 0.15 & 2.19 $\pm$ 0.77 & $>$0.48\\
W0147--0923 & 9.20 $\pm$ 0.35 & 9.02 $\pm$ 7.25 & $>$1.44\\
W0226+0514 & 8.95 $\pm$ 0.28 & 8.08 $\pm$ 5.24 & $>$0.71\\
W1103--1826 & 9.18 $\pm$ 0.30 & 3.60 $\pm$ 2.49 & $>$0.54\\
W1136+4236 & $<$9.96 $\pm$ 0.13 & $>$0.28 $\pm$ 0.09 & $>$0.26\\
W1719+0446 & $<$10.01 $\pm$ 0.46 & $>$0.81 $\pm$ 0.86 & $>$0.83\\
W2026+0716 & $<$10.01 $\pm$ 0.27 & $>$0.43 $\pm$ 0.27 & $>$0.45\\
W2136--1631 & 8.08 $\pm$ 0.07 & 15.98 $\pm$ 2.40 & $>$0.19\\
W2216+0723 & $<$10.02 $\pm$ 0.17 & $>$0.23 $\pm$ 0.09 & $>$0.24
\enddata
\tablecomments{$M_{\rm BH}$ is the mass assuming the local $M_{\rm BH}$--$\sigma_{*}$ relation and using $\sigma_{\rm narrow}$ as a substitute for $\sigma_{*}$ from [\ion{O}{2}], H$\beta$, or H$\alpha$ lines with S/N$\ge$5. $f_{\rm Edd}$ is the Eddington ratio, $L_{\rm bol}/L_{\rm Edd}$ where $L_{\rm bol}$ is derived from the SED fit (e.g., Figure 1), $L_{\rm Edd}$ is the Eddington luminosity. $f_{\rm Edd}$ values are derived using $\log M_{\rm BH}/M_{\odot}=8.49+4.377\log\{\rm \sigma_{*}\, (\kms)/200\}$ (\citealt{Kor13} and a 1-1 relation between $\sigma_{*}$ and $\sigma_{\rm narrow}$, e.g., \citealt{Ben18}), and an observed upper limit of $M_{\rm BH}\sim10^{10}M_{\odot}$ for the most massive SDSS quasars \citep{Jun17}.} 
\end{deluxetable} 
\endgroup

To better tell whether the super-Eddington accretion seen in some of the Hot DOGs is real or due to systematic overestimation, we estimate the $f_{\rm Edd}$ values again from considering an upper limit of $M_{\rm BH}\sim10^{10} M_{\odot}$ for the most massive $z\sim2$ quasars \citep{Jun17}. Listed in Table 10, the corresponding lower limits on $f_{\rm Edd}$ values are around 0.2--1.4 (mean 0.6), indicating near Eddington-limited accretion even if $M_{\rm BH}$ reaches $\sim10^{10} M_{\odot}$, and super-Eddington accretion if $M_{\rm BH}\lesssim 5 \times 10^{9} M_{\odot}$. In any case, we find without any measurement of $M_{\rm BH}$ that Hot DOGs are likely accreting near or beyond the Eddington limit. If we trust the $f_{\rm Edd}$ values from the $M_{\rm BH}$--$\sigma_{*}$ relation however, 4/9 sources have $f_{\rm Edd}$ exceeding 3 (but with large measurement uncertainties), which is hard to explain with models allowing accretion modes up to several times the Eddington limit (e.g., see discussion in \citealt{Tsa18}). We thus find it likely that some of our $f_{\rm Edd}$ values are largely overestimated either due to an underestimated $M_{\rm BH}$, or an overestimated $L_{\rm bol}$.

Under the merger-driven AGN triggering mechanism (e.g., \citealt{Hop08}; \citealt{Hic09}), starbursts in the merging system will be followed by obscured BH growth. Using a simple BH-galaxy mass scaling relation, $M_{\rm BH}$ values for Hot DOGs are more likely higher relative to the central host mass due to the time delay between BH and galaxy growth (e.g., \citealt{Urr12}), and the bulge formation lagging the BH growth (e.g., \citealt{Pen06}; \citealt{Jun17}) at $z\sim2$. Alternative explanations for extremely high $f_{\rm Edd}$ values for Hot DOGs could be due to $L_{\rm bol}$ being overestimated by anisotropic radiation, (e.g., \citealt{Abr88}; \citealt{Wan14}), but this is less likely significant as the SED peaking in the infrared is relatively isotropic. Gravitational lensing is also a less likely explanation given the morphologies of Hot DOGs imaged by the {\it Hubble Space Telescope} (e.g., \citealt{Tsa15}). 

To summarize, our sample of Hot DOGs show high $f_{\rm Edd}$ values, with lower limits often near the Eddington limit. This may be produced by underestimated $M_{\rm BH}$ ($\sigma_{*}$) values, but with large potential uncertainties in both the $M_{\rm BH}$ and $L_{\rm bol}$, it is hard to tell from our data whether super-Eddington accretion is occurring. The efficient accretion probed by our objects is consistent with scenarios with BH feeding in luminous, obscured AGN triggering strong ionized gas outflows. Our sample is also in line with the more X-ray luminous population of SMGs showing a stronger, near-Eddington limited accretion than the X-ray weaker counterpart \citep{Ale08}, following the evolutionary model predictions. However, it is in tension with the absence of hard X-ray selected local AGN population with high $f_{\rm Edd}$ and $E(B-V)$ \citep{Ric17c}, suggesting that the local population might simply lack the rapidly accreting, obscured AGN seen at $z\sim2$ (Jun et al., in preparation).

\section{Summary}
We classified the rest-frame UV--optical spectra of 12 Hot DOGs at $z\sim2$, and quantified the ionized gas outflows in [\ion{O}{3}]. The main results are summarized as below.

1. The redshifts for Hot DOGs in this sample peak around 2--3 and are consistent with the distribution obtained from rest-frame UV spectroscopy.
Using a wide spectral window in this study, we found 2 mismatches out of 8 previous redshift estimates that were based on narrower spectral coverage.

2. 7/8 Hot DOGs are classified as AGN in the BPT diagram, confirming the expectation from the strong mid-IR peak in their SEDs. The Seyfert types are 1.8 or higher for 6/8 objects, indicating the broad line regions are mostly obscured. The [\ion{O}{2}]/[\ion{O}{3}] ratios, which potentially traces the relative contribution of star-formation over AGN activity, vary between 0.1--2.6 and are lower for AGN exclusively in all three BPT diagrams.

3. Extinction within the narrow line region estimated from the Balmer decrement indicates the $E(B-V)$ values around 0.3--0.7 mag. This is about an order of magnitude smaller than the values from SED modeling, indicating most of the obscuration is concentrated interior to the narrow line region. As the narrow line region is extended up to kpc scales, however, we infer that obscuration exists throughout the host galaxy. $E(B-V)$ values from EW$_{[\rm O\,III]}$ independently tracing the obscuration toward the AGN center have lower limits of 0--1.1 mag.

4. Ionized gas outflows are seen in 8/9 [\ion{O}{3}] lines with S/N$\ge$5, broad enough ($\sigma_{[\rm O\,III], broad}>400\kms$) to be distinguished from gravitational motion in the most massive bulges. The median broadening and blueshift of the broad [\ion{O}{3}] component are 1100 \kms\ and $-$1100 \kms\ respectively, much stronger than lower luminosity quasars from the literature. The outflow kinematics for Hot DOGs together with other highly obscured AGN, are contrary to the the simplest orientation-driven models of type 2 AGN. Comparing with the literature, the luminosity dependence on outflow kinematics is greater than that of AGN type, favoring a physical and evolutionary origin for producing ionized gas outflows. 

Normalized, median outflow quantities are $\dot{M}_{\rm out}/\dot{M}_{\rm acc}\sim$5.7, $\dot{E}_{\rm out}/L_{\rm bol}\sim$0.32\%, and $\dot{P}_{\rm out}/(L_{\rm bol}/c)\sim$0.60, all corrected for extinction but having a large scatter and uncertainty ($\sim$an order of magnitude each). Mass outflow rates are comparable to or higher than star formation or mass accretion rates, indicating outflows in Hot DOGs may be efficient depleting gas, whereas the energy injection rate and momentum flux are sufficient to quench the star formation in galaxy simulations.

5. We find further hints of outflow signatures in S/N$\ge$5 H$\alpha$ lines (5/7), showing a broad ($\sigma_{\rm broad/narrow}>400\kms$) component blueshifted by 0.5--1.2 times that of the [\ion{O}{3}]. Even some S/N$\ge$5 [\ion{O}{2}] lines (3/6) have $\sigma_{\rm broad/narrow}>400\kms$, supporting the presence of ionized gas outflows in multiple ionization states. The signs for outflows in the Balmer lines and the relatively higher obscuration towards the broad line region than the narrow line region can complicate $M_{\rm BH}$ estimation for luminous, obscured AGN. 

6. Alternative estimates of the Eddington ratio based on the $M_{\rm BH}$--$\sigma_{\rm narrow}$ relation and the observed upper limit on the $M_{\rm BH}$ values for AGN consistently suggest near-Eddington or even stronger accretion, supporting that luminous, obscured AGN activity from vast amounts of gas fed into the BH is responsible for producing feedback.
\\\\
We summarize several remaining issues on outflows from luminous, obscured AGN. We note that ionized gas represents only a fraction of the total gas, and our outflow quantities are commonly underestimated. Observations of multi-phase outflows from gas spatially resolved and distributed in various locations of the AGN (e.g., \citealt{Dia16,Dia18}, \citealt{Fio17}), will help understand the dominant phase and location of the energy output injected into the ISM. As we find a high fraction and strength of ionized gas outflows among the most luminous, obscured AGN, the demographics of the merger-triggered AGN activity along various observed samples should eventually be analyzed to resolve the issue whether the scarcity of the most massive galaxies matches the energetics of AGN feedback quenching star formation.

\acknowledgments
We thank the anonymous referee for the valuable comments improving the paper, and the ESO staff for kindly observing and providing processed data products. This research was supported by Basic Science Research Program through the National Research Foundation of Korea (NRF) funded by the Ministry of Education (NRF-2017R1A6A3A04005158). RJA was supported by FONDECYT grant number 1191124. FEB acknowledges support from CONICYT-Chile (Basal AFB-170002) and the Ministry of Economy, Development, and Tourism's Millennium Science Initiative through grant IC120009, awarded to The Millennium Institute of Astrophysics, MAS. JW acknowledges support from MSTC through grant 2016YFA0400702 and NSFC 11673029.
This publication makes use of data products from the {\it Wide-field Infrared Survey Explorer}, which is a joint project of the University of California, Los Angeles, and the Jet Propulsion Laboratory/California Institute of Technology, funded by the National Aeronautics and Space Administration.


\begin{thebibliography}{}
\bibitem[Abramowicz et al.(1988)]{Abr88} Abramowicz, M.~A., Czerny, B., Lasota, J.~P., et al.\ 1988, \apj, 332, 646
\bibitem[Alexander et al.(2008)]{Ale08} Alexander, D.~M., Brandt, W.~N., Smail, I., et al.\ 2008, \aj, 135, 1968
\bibitem[Antonucci(1993)]{Ant93} Antonucci, R.\ 1993, \araa, 31, 473 
\bibitem[Asmus et al.(2016)]{Asm16} Asmus, D., H{\"o}nig, S.~F., \& Gandhi, P.\ 2016, \apj, 822, 109 
\bibitem[Assef et al.(2015)]{Ass15} Assef, R.~J., Eisenhardt, P.~R.~M., Stern, D., et al.\ 2015, \apj, 804, 27 
\bibitem[Assef et al.(2016)]{Ass16} Assef, R.~J., Walton, D.~J., Brightman, M., et al.\ 2016, \apj, 819, 111
\bibitem[Assef et al.(2019)]{Ass19} Assef, R.~J., Brightman, M., Walton, D.~J., et al.\ 2019, arXiv e-prints, arXiv:1905.04320
\bibitem[Bae \& Woo(2014)]{Bae14} Bae, H.-J., \& Woo, J.-H.\ 2014, \apj, 795, 30
\bibitem[Bae \& Woo(2016)]{Bae16} Bae, H.-J., \& Woo, J.-H.\ 2016, \apj, 828, 97  
\bibitem[Bae et al.(2017)]{Bae17} Bae, H.-J., Woo, J.-H., Karouzos, M., et al.\ 2017, \apj, 837, 91 
\bibitem[Baldwin(1977)]{Bal77} Baldwin, J.~A.\ 1977, \apj, 214, 679 
\bibitem[Baldwin et al.(1981)]{Bal81} Baldwin, J.~A., Phillips, M.~M., \& Terlevich, R.\ 1981, \pasp, 93, 5 
\bibitem[Banerji et al.(2015)]{Ban15} Banerji, M., Alaghband-Zadeh, S., Hewett, P.~C., \& McMahon, R.~G.\ 2015, \mnras, 447, 3368 
\bibitem[Banerji et al.(2012)]{Ban12} Banerji, M., McMahon, R.~G., Hewett, P.~C., et al.\ 2012, \mnras, 427, 2275 
\bibitem[Banerji et al.(2013)]{Ban13} Banerji, M., McMahon, R.~G., Hewett, P.~C., Gonzalez-Solares, E., \& Koposov, S.~E.\ 2013, \mnras, 429, L55 
\bibitem[Baron \& Netzer(2019)]{Bar19} Baron, D., \& Netzer, H.\ 2019, \mnras, 486, 4290
\bibitem[Bennert et al.(2018)]{Ben18} Bennert, V.~N., Loveland, D., Donohue, E., et al.\ 2018, \mnras, 481, 138 
\bibitem[Bischetti et al.(2017)]{Bis17} Bischetti, M., Piconcelli, E., Vietri, G., et al.\ 2017, \aap, 598, A122
\bibitem[Blain et al.(2002)]{Bla02} Blain, A.~W., Smail, I., Ivison, R.~J., et al.\ 2002, \physrep, 369, 111
\bibitem[Bridge et al.(2013)]{Bri13} Bridge, C.~R., Blain, A., Borys, C.~J.~K., et al.\ 2013, \apj, 769, 91 
\bibitem[Brotherton(1996)]{Bro96} Brotherton, M.~S.\ 1996, \apjs, 102, 1 
\bibitem[Brusa et al.(2015)]{Bru15} Brusa, M., Bongiorno, A., Cresci, G., et al.\ 2015, \mnras, 446, 2394
\bibitem[Buchner \& Bauer(2017)]{Buc17} Buchner, J., \& Bauer, F.~E.\ 2017, \mnras, 465, 4348 
\bibitem[Cano-D{\'\i}az et al.(2012)]{Can12} Cano-D{\'\i}az, M., Maiolino, R., Marconi, A., et al.\ 2012, \aap, 537, L8
\bibitem[Carniani et al.(2015)]{Car15} Carniani, S., Marconi, A., Maiolino, R., et al.\ 2015, \aap, 580, A102
\bibitem[Chapman et al.(2005)]{Cha05} Chapman, S.~C., Blain, A.~W., Smail, I., \& Ivison, R.~J.\ 2005, \apj, 622, 772 
\bibitem[Choi et al.(2012)]{Cho12} Choi, E., Ostriker, J.~P., Naab, T., \& Johansson, P.~H.\ 2012, \apj, 754, 125 
\bibitem[Crenshaw \& Kraemer(2000)]{Cre00} Crenshaw, D.~M., \& Kraemer, S.~B.\ 2000, \apjl, 532, L101  
\bibitem[Crenshaw et al.(2010)]{Cre10} Crenshaw, D.~M., Schmitt, H.~R., Kraemer, S.~B., Mushotzky, R.~F., \& Dunn, J.~P.\ 2010, \apj, 708, 419 
\bibitem[Cresci et al.(2015)]{Cre15} Cresci, G., Mainieri, V., Brusa, M., et al.\ 2015, \apj, 799, 82
\bibitem[Croton et al.(2006)]{Cro06} Croton, D.~J., Springel, V., White, S.~D.~M., et al.\ 2006, \mnras, 365, 11
\bibitem[Dey et al.(2008)]{Dey08} Dey, A., Soifer, B.~T., Desai, V., et al.\ 2008, \apj, 677, 943 
\bibitem[D{\'\i}az-Santos et al.(2016)]{Dia16} D{\'\i}az-Santos, T., Assef, R.~J., Blain, A.~W., et al.\ 2016, \apjl, 816, L6 
\bibitem[D{\'\i}az-Santos et al.(2018)]{Dia18} D{\'\i}az-Santos, T., Assef, R.~J., Blain, A.~W., et al.\ 2018, Science, 362, 1034
\bibitem[Di Matteo et al.(2005)]{Dim05} Di Matteo, T., Springel, V., \& Hernquist, L.\ 2005, \nat, 433, 604
\bibitem[DiPompeo et al.(2018)]{Dip18} DiPompeo, M.~A., Hickox, R.~C., Carroll, C.~M., et al.\ 2018, \apj, 856, 76
\bibitem[Dimitrijevi{\'c} et al.(2007)]{Dim07} Dimitrijevi{\'c}, M.~S., Popovi{\'c}, L.~{\v C}., Kova{\v c}evi{\'c}, J., Da{\v c}i{\'c}, M., \& Ili{\'c}, D.\ 2007, \mnras, 374, 1181 
\bibitem[Dubois et al.(2014)]{Dub14} Dubois, Y., Pichon, C., Welker, C., et al.\ 2014, \mnras, 444, 1453 
\bibitem[Eisenhardt et al.(2012)]{Eis12} Eisenhardt, P.~R.~M., Wu, J., Tsai, C.-W., et al.\ 2012, \apj, 755, 173 
\bibitem[Fabian(2012)]{Fab12} Fabian, A.~C.\ 2012, \araa, 50, 455
\bibitem[Fan et al.(2016)]{Fan16} Fan, L., Han, Y., Fang, G., et al.\ 2016, \apjl, 822, L32 
\bibitem[Farrah et al.(2017)]{Far17} Farrah, D., Petty, S., Connolly, B., et al.\ 2017, \apj, 844, 106
\bibitem[Ferland \& Osterbrock(1986)]{Fer86} Ferland, G.~J., \& Osterbrock, D.~E.\ 1986, \apj, 300, 658
\bibitem[Feruglio et al.(2015)]{Fer15} Feruglio, C., Fiore, F., Carniani, S., et al.\ 2015, \aap, 583, A99
\bibitem[Fiore et al.(2017)]{Fio17} Fiore, F., Feruglio, C., Shankar, F., et al.\ 2017, \aap, 601, A143  
\bibitem[Fischer et al.(2014)]{Fis14} Fischer, T.~C., Crenshaw, D.~M., Kraemer, S.~B., et al.\ 2014, \apj, 785, 25
\bibitem[Kim \& Im(2018)]{Kim18} Kim, D., \& Im, M.\ 2018, \aap, 610, A31
\bibitem[King(2003)]{Kin03} King, A.\ 2003, \apjl, 596, L27 
\bibitem[Freudling et al.(2013)]{Fre13} Freudling, W., Romaniello, M., Bramich, D.~M., et al.\ 2013, \aap, 559, A96  
\bibitem[Glikman et al.(2018)]{Gli18} Glikman, E., Lacy, M., LaMassa, S., et al.\ 2018, \apj, 861, 37
\bibitem[Glikman et al.(2012)]{Gli12} Glikman, E., Urrutia, T., Lacy, M., et al.\ 2012, \apj, 757, 51 
\bibitem[Gobat et al.(2018)]{Gob18} Gobat, R., Daddi, E., Magdis, G., et al.\ 2018, Nature Astronomy, 2, 239
\bibitem[Greene et al.(2014)]{Gre14} Greene, J.~E., Alexandroff, R., Strauss, M.~A., et al.\ 2014, \apj, 788, 91
\bibitem[Greene \& Ho(2005a)]{Gre05a} Greene, J.~E., \& Ho, L.~C.\ 2005a, \apj, 627, 721 
\bibitem[Greene \& Ho(2005b)]{Gre05b} Greene, J.~E., \& Ho, L.~C.\ 2005b, \apj, 630, 122
\bibitem[Hainline et al.(2014)]{Hai14} Hainline, K.~N., Hickox, R.~C., Greene, J.~E., et al.\ 2014, \apj, 787, 65 
\bibitem[Hamann et al.(2017)]{Ham17} Hamann, F., Zakamska, N.~L., Ross, N., et al.\ 2017, \mnras, 464, 3431 
\bibitem[Harrison et al.(2014)]{Har14} Harrison, C.~M., Alexander, D.~M., Mullaney, J.~R., \& Swinbank, A.~M.\ 2014, \mnras, 441, 3306 
\bibitem[Harrison et al.(2016)]{Har16} Harrison, C.~M., Alexander, D.~M., Mullaney, J.~R., et al.\ 2016, \mnras, 456, 1195
\bibitem[Harrison et al.(2018)]{Har18} Harrison, C.~M., Costa, T., Tadhunter, C.~N., et al.\ 2018, Nature Astronomy, 2, 198
\bibitem[Heckman et al.(2004)]{Hec04} Heckman, T.~M., Kauffmann, G., Brinchmann, J., et al.\ 2004, \apj, 613, 109 
\bibitem[Hickox et al.(2009)]{Hic09} Hickox, R.~C., Jones, C., Forman, W.~R., et al.\ 2009, \apj, 696, 891 
\bibitem[Ho(2005)]{Ho05} Ho, L.~C.\ 2005, \apj, 629, 680 
\bibitem[Holt et al.(2006)]{Hol06} Holt, J., Tadhunter, C., Morganti, R., et al.\ 2006, \mnras, 370, 1633 
\bibitem[Hopkins \& Elvis(2010)]{Hop10} Hopkins, P.~F., \& Elvis, M.\ 2010, \mnras, 401, 7 
\bibitem[Hopkins et al.(2008)]{Hop08} Hopkins, P.~F., Hernquist, L., Cox, T.~J., \& Kere{\v s}, D.\ 2008, \apjs, 175, 356
\bibitem[H{\"o}nig \& Beckert(2007)]{Hon07} H{\"o}nig, S.~F., \& Beckert, T.\ 2007, \mnras, 380, 1172 
\bibitem[H{\"o}nig et al.(2013)]{Hon13} H{\"o}nig, S.~F., Kishimoto, M., Tristram, K.~R.~W., et al.\ 2013, \apj, 771, 87
\bibitem[Husemann et al.(2013)]{Hus13} Husemann, B., Wisotzki, L., S{\'a}nchez, S.~F., \& Jahnke, K.\ 2013, \aap, 549, A43
\bibitem[Im et al.(1997)]{Im97} Im, M., Griffiths, R. E., \& Ratnatunga, K. U.\ 1997, \apj, 475, 457
\bibitem[Jones et al.(2014)]{Jon14} Jones, S.~F., Blain, A.~W., Stern, D., et al.\ 2014, \mnras, 443, 146  
\bibitem[Jun \& Im(2013)]{Jun13} Jun, H.~D., \& Im, M.\ 2013, \apj, 779, 104
\bibitem[Jun et al.(2017)]{Jun17} Jun, H.~D., Im, M., Kim, D., \& Stern, D.\ 2017, \apj, 838, 41 
\bibitem[Jun et al.(2015)]{Jun15} Jun, H.~D., Im, M., Lee, H.~M., et al.\ 2015, \apj, 806, 109 
\bibitem[Kang \& Woo(2018)]{Kan18} Kang, D., \& Woo, J.-H.\ 2018, \apj, 864, 124
\bibitem[Kang et al.(2017)]{Kan17} Kang, D., Woo, J.-H., \& Bae, H.-J.\ 2017, \apj, 845, 131 
\bibitem[Karouzos et al.(2016)]{Kar16} Karouzos, M., Woo, J.-H., \& Bae, H.-J.\ 2016, \apj, 833, 171 
\bibitem[Kauffmann et al.(2003)]{Kau03} Kauffmann, G., Heckman, T.~M., Tremonti, C., et al.\ 2003, \mnras, 346, 1055 
\bibitem[Kausch et al.(2015)]{Kau15} Kausch, W., Noll, S., Smette, A., et al.\ 2015, \aap, 576, A78 
\bibitem[Kewley et al.(2001)]{Kew01} Kewley, L.~J., Dopita, M.~A., Sutherland, R.~S., Heisler, C.~A., \& Trevena, J.\ 2001, \apj, 556, 121 
\bibitem[Kewley et al.(2006)]{Kew06} Kewley, L.~J., Groves, B., Kauffmann, G., \& Heckman, T.\ 2006, \mnras, 372, 961 
\bibitem[Kim et al.(2006)]{Kim06} Kim, M., Ho, L.~C., \& Im, M.\ 2006, \apj, 642, 702 
\bibitem[Kim et al.(2008)]{Kim08} Kim, M., Ho, L.~C., Peng, C.~Y., et al.\ 2008, \apj, 687, 767
\bibitem[Kormendy \& Ho(2013)]{Kor13} Kormendy, J., \& Ho, L.~C.\ 2013, \araa, 51, 511
\bibitem[Lacy et al.(2015)]{Lac15} Lacy, M., Ridgway, S.~E., Sajina, A., et al.\ 2015, \apj, 802, 102 
\bibitem[Leung et al.(2017)]{Leu17} Leung, G.~C.~K., Coil, A.~L., Azadi, M., et al.\ 2017, \apj, 849, 48
\bibitem[Liu et al.(2013)]{Liu13} Liu, G., Zakamska, N.~L., Greene, J.~E., Nesvadba, N.~P.~H., \& Liu, X.\ 2013, \mnras, 436, 2576 
\bibitem[Liu et al.(2014)]{Liu14} Liu, G., Zakamska, N.~L., \& Greene, J.~E.\ 2014, \mnras, 442, 1303 
\bibitem[Maddox(2018)]{Mad18} Maddox, N.\ 2018, \mnras, 480, 5203 
\bibitem[Magnelli et al.(2012)]{Mag12} Magnelli, B., Lutz, D., Santini, P., et al.\ 2012, \aap, 539, A155 
\bibitem[Maiolino et al.(2012)]{Mai12} Maiolino, R., Gallerani, S., Neri, R., et al.\ 2012, \mnras, 425, L66 
\bibitem[Marin(2016)]{Mar16} Marin, F.\ 2016, \mnras, 460, 3679
\bibitem[Markwardt(2009)]{Mar09}Markwardt, C. B. 2009, in ASP Conf. Ser. 411, Astronomical Data Analysis Software and Systems XVIII, ed. D. A. Bohlender, D. Durand, \& P. Dowler (San Francisco, CA: ASP), 251
\bibitem[Mullaney et al.(2013)]{Mul13} Mullaney, J.~R., Alexander, D.~M., Fine, S., et al.\ 2013, \mnras, 433, 622 
\bibitem[Murakami et al.(2007)]{Mur07} Murakami, H., Baba, H., Barthel, P., et al.\ 2007, \pasj, 59, 369
\bibitem[Narayanan et al.(2010)]{Nar10} Narayanan, D., Dey, A., Hayward, C.~C., et al.\ 2010, \mnras, 407, 1701
\bibitem[Nenkova et al.(2008)]{Nen08} Nenkova, M., Sirocky, M.~M., Nikutta, R., Ivezi{\'c}, {\v Z}., \& Elitzur, M.\ 2008, \apj, 685, 160 
\bibitem[Neugebauer et al.(1984)]{Neu84} Neugebauer, G., Habing, H.~J., van Duinen, R., et al.\ 1984, \apjl, 278, L1 
\bibitem[Nesvadba et al.(2006)]{Nes06} Nesvadba, N.~P.~H., Lehnert, M.~D., Eisenhauer, F., et al.\ 2006, \apj, 650, 693 
\bibitem[Nesvadba et al.(2011)]{Nes11} Nesvadba, N.~P.~H., Polletta, M., Lehnert, M.~D., et al.\ 2011, \mnras, 415, 2359 
\bibitem[Osterbrock(1981)]{Ost81} Osterbrock, D.~E.\ 1981, \apj, 249, 462 
\bibitem[Osterbrock(1989)]{Ost89} Osterbrock, D.~E.\ 1989, Astrophysics of Gaseous Nebulae and Active Galactic Nuclei (Mill Valley, CA: Univ. Science Books)
\bibitem[Peng et al.(2006)]{Pen06} Peng, C.~Y., Impey, C.~D., Rix, H.-W., et al.\ 2006, \apj, 649, 616
\bibitem[Perna et al.(2015)]{Per15} Perna, M., Brusa, M., Cresci, G., et al.\ 2015, \aap, 574, A82 
\bibitem[Perrotta et al.(2019)]{Per19} Perrotta, S., Hamann, F., Zakamska, N.~L., et al.\ 2019, \mnras, 488, 4126
\bibitem[Piconcelli et al.(2015)]{Pic15} Piconcelli, E., Vignali, C., Bianchi, S., et al.\ 2015, \aap, 574, L9 
\bibitem[Polletta et al.(2008)]{Pol08} Polletta, M., Weedman, D., H{\"o}nig, S., et al.\ 2008, \apj, 675, 960
\bibitem[Rakshit et al.(2017)]{Rak17} Rakshit, S., Stalin, C.~S., Chand, H., \& Zhang, X.-G.\ 2017, \apjs, 229, 39 
\bibitem[Rakshit \& Woo(2018)]{Rak18} Rakshit, S., \& Woo, J.-H.\ 2018, \apj, 865, 5  
\bibitem[Ricci et al.(2017a)]{Ric17a} Ricci, C., Assef, R.~J., Stern, D., et al.\ 2017a, \apj, 835, 105 
\bibitem[Ricci et al.(2017b)]{Ric17b} Ricci, C., Bauer, F.~E., Treister, E., et al.\ 2017b, \mnras, 468, 1273 
\bibitem[Ricci et al.(2017c)]{Ric17c} Ricci, C., Trakhtenbrot, B., Koss, M.~J., et al.\ 2017c, \nat, 549, 488 
\bibitem[Richards et al.(2006)]{Ric06} Richards, G.~T., Lacy, M., Storrie-Lombardi, L.~J., et al.\ 2006, \apjs, 166, 470
\bibitem[Rigby et al.(2006)]{Rig06} Rigby, J.~R., Rieke, G.~H., Donley, J.~L., Alonso-Herrero, A., \& P{\'e}rez-Gonz{\'a}lez, P.~G.\ 2006, \apj, 645, 115 
\bibitem[Rodr{\'\i}guez Zaur{\'\i}n et al.(2013)]{Rod13} Rodr{\'\i}guez Zaur{\'\i}n, J., Tadhunter, C.~N., Rose, M., et al.\ 2013, \mnras, 432, 138
\bibitem[Ross et al.(2015)]{Ros15} Ross, N.~P., Hamann, F., Zakamska, N.~L., et al.\ 2015, \mnras, 453, 3932 
\bibitem[Roth et al.(2012)]{Rot12} Roth, N., Kasen, D., Hopkins, P.~F., \& Quataert, E.\ 2012, \apj, 759, 36 
\bibitem[Sanders \& Mirabel(1996)]{San96} Sanders, D.~B., \& Mirabel, I.~F.\ 1996, \araa, 34, 749 
\bibitem[Sanders et al.(1988)]{San88} Sanders, D.~B., Soifer, B.~T., Elias, J.~H., et al.\ 1988, \apj, 325, 74 
\bibitem[Schaye et al.(2015)]{Sch15} Schaye, J., Crain, R.~A., Bower, R.~G., et al.\ 2015, \mnras, 446, 521
\bibitem[Schlafly \& Finkbeiner(2011)]{Sch11} Schlafly, E.~F., \& Finkbeiner, D.~P.\ 2011, \apj, 737, 103 
\bibitem[Sexton et al.(2019)]{Sex19} Sexton, R.~O., Canalizo, G., Hiner, K.~D., et al.\ 2019, \apj, 878, 101
\bibitem[Shen \& Liu(2012)]{She12} Shen, Y., \& Liu, X.\ 2012, \apj, 753, 125 
\bibitem[Shen et al.(2011)]{She11} Shen, Y., Richards, G.~T., Strauss, M.~A., et al.\ 2011, \apjs, 194, 45 
\bibitem[Shen(2016)]{She16} Shen, Y.\ 2016, \apj, 817, 55 
\bibitem[Shen \& Ho(2014)]{She14} Shen, Y., \& Ho, L.~C.\ 2014, \nat, 513, 210
\bibitem[Silk \& Rees(1998)]{Sil98} Silk, J., \& Rees, M.~J.\ 1998, \aap, 331, L1 
\bibitem[Smette et al.(2015)]{Sme15} Smette, A., Sana, H., Noll, S., et al.\ 2015, \aap, 576, A77 
\bibitem[Soltan(1982)]{Sol82} Soltan, A.\ 1982, \mnras, 200, 115 
\bibitem[Stern et al.(2014)]{Ste14} Stern, D., Lansbury, G.~B., Assef, R.~J., et al.\ 2014, \apj, 794, 102 
\bibitem[Storchi-Bergmann et al.(2018)]{Sto18} Storchi-Bergmann, T., Dall'Agnol de Oliveira, B., Longo Micchi, L.~F., et al.\ 2018, \apj, 868, 14 
\bibitem[Storey \& Zeippen(2000)]{Sto00} Storey, P.~J., \& Zeippen, C.~J.\ 2000, \mnras, 312, 813 
\bibitem[Temple et al.(2019)]{Tem19} Temple, M.~J., Banerji, M., Hewett, P.~C., et al.\ 2019, \mnras, 487, 2594
\bibitem[Toba et al.(2017)]{Tob17} Toba, Y., Bae, H.-J., Nagao, T., et al.\ 2017, \apj, 850, 140 
\bibitem[Tombesi et al.(2015)]{Tom15} Tombesi, F., Mel{\'e}ndez, M., Veilleux, S., et al.\ 2015, \nat, 519, 436 
\bibitem[Tsai et al.(2018)]{Tsa18} Tsai, C.-W., Eisenhardt, P.~R.~M., Jun, H.~D., et al.\ 2018, \apj, 868, 15 
\bibitem[Tsai et al.(2015)]{Tsa15} Tsai, C.-W., Eisenhardt, P.~R.~M., Wu, J., et al.\ 2015, \apj, 805, 90 
\bibitem[Urrutia et al.(2012)]{Urr12} Urrutia, T., Lacy, M., Spoon, H., et al.\ 2012, \apj, 757, 125 
\bibitem[Urry \& Padovani(1995)]{Urr95} Urry, C., \& Padovani, P. 1995, \pasp, 107, 803
\bibitem[Veilleux et al.(1994)]{Vei94} Veilleux, S., Cecil, G., Bland-Hawthorn, J., et al.\ 1994, \apj, 433, 48 
\bibitem[Vietri et al.(2018)]{Vie18} Vietri, G., Piconcelli, E., Bischetti, M., et al.\ 2018, \aap, 617, A81
\bibitem[Villar-Mart{\'{\i}}n et al.(2011)]{Vil11} Villar-Mart{\'{\i}}n, M., Humphrey, A., Delgado, R.~G., Colina, L., \& Arribas, S.\ 2011, \mnras, 418, 2032 
\bibitem[Vernet et al.(2011)]{Ver11} Vernet, J., Dekker, H., D'Odorico, S., et al.\ 2011, \aap, 536, A105 
\bibitem[Wang et al.(2014)]{Wan14} Wang, J.-M., Qiu, J., Du, P., et al.\ 2014, \apj, 797, 65
\bibitem[Weinberger et al.(2017)]{Wei17} Weinberger, R., Springel, V., Hernquist, L., et al.\ 2017, \mnras, 465, 3291
\bibitem[Werner et al.(2004)]{Wer04} Werner, M.~W., Roellig, T.~L., Low, F.~J., et al.\ 2004, \apjs, 154, 1 
\bibitem[Winkler(1992)]{Win92} Winkler, H.\ 1992, \mnras, 257, 677 
\bibitem[Woo et al.(2016)]{Woo16} Woo, J.-H., Bae, H.-J., Son, D., \& Karouzos, M.\ 2016, \apj, 817, 108 
\bibitem[Woo et al.(2013)]{Woo13} Woo, J.-H., Schulze, A., Park, D., et al.\ 2013, \apj, 772, 49 
\bibitem[Wright et al.(2010)]{Wri10} Wright, E.~L., Eisenhardt, P.~R.~M., Mainzer, A.~K., et al.\ 2010, \aj, 140, 1868
\bibitem[Wu et al.(2018)]{Wu18} Wu, J., Jun, H.~D., Assef, R.~J., et al.\ 2018, \apj, 852, 96
\bibitem[Wu et al.(2012)]{Wu12} Wu, J., Tsai, C.-W., Sayers, J., et al.\ 2012, \apj, 756, 96 
\bibitem[Yan et al.(2006)]{Yan06} Yan, R., Newman, J.~A., Faber, S.~M., et al.\ 2006, \apj, 648, 281 
\bibitem[Zakamska \& Greene(2014)]{Zak14} Zakamska, N.~L., \& Greene, J.~E.\ 2014, \mnras, 442, 784 
\bibitem[Zakamska et al.(2016)]{Zak16} Zakamska, N.~L., Hamann, F., P{\^a}ris, I., et al.\ 2016, \mnras, 459, 3144 
\bibitem[Zakamska et al.(2005)]{Zak05} Zakamska, N.~L., Schmidt, G.~D., Smith, P.~S., et al.\ 2005, \aj, 129, 1212 
\bibitem[Zakamska et al.(2003)]{Zak03} Zakamska, N.~L., Strauss, M.~A., Krolik, J.~H., et al.\ 2003, \aj, 126, 2125 
\end{thebibliography}
\end{document}